\documentclass[11pt,a4paper]{article}

\usepackage[UKenglish]{babel}
\usepackage[T1]{fontenc}
\usepackage{lmodern}
\usepackage{microtype}
\usepackage{geometry}
\usepackage{authblk}

\usepackage{amsmath,amssymb,amsthm,mathtools}
\usepackage{booktabs}
\usepackage{float}
\usepackage{array}
\usepackage{bm}
\usepackage{csquotes}
\usepackage[
  backend=biber,
  style=numeric-comp,
  sorting=none,
  giveninits=true,
  maxbibnames=99,
  doi=false,
  url=false,
  isbn=false,
  eprint=false
]{biblatex}
\DeclareFieldFormat[article,inproceedings,incollection]
\renewbibmacro{in:}{}
\AtEveryBibitem{\clearfield{month}\clearfield{note}}
\usepackage[hidelinks]{hyperref}
\usepackage[nameinlink,capitalise,noabbrev]{cleveref}
\usepackage{titlesec}
\titlespacing*{\section}{0pt}{1.5ex plus .4ex minus .2ex}{0.8ex}
\titlespacing*{\subsection}{0pt}{1.1ex plus .3ex minus .2ex}{0.5ex}
\titlespacing*{\subsubsection}{0pt}{0.9ex plus .2ex minus .2ex}{0.4ex}

\allowdisplaybreaks

\newtheorem{theorem}{Theorem}[section]
\newtheorem{lemma}[theorem]{Lemma}

\newtheorem{corollary}[theorem]{Corollary}
\newtheorem{observation}[theorem]{Observation}

\crefname{theorem}{Theorem}{Theorems}
\crefname{lemma}{Lemma}{Lemmas}
\crefname{proposition}{Proposition}{Propositions}
\crefname{corollary}{Corollary}{Corollaries}
\crefname{observation}{Observation}{Observations}
\crefname{section}{Section}{Sections}
\crefname{subsection}{Section}{Sections}
\crefname{subsubsection}{Section}{Sections}

\newcommand{\E}{\mathbb{E}}
\newcommand{\Pp}{\mathbb{P}}
\newcommand{\R}{\mathbb{R}}
\newcommand{\F}{\mathcal{F}}
\newcommand{\1}{\mathbf{1}}
\newcommand{\Var}{\operatorname{Var}}
\newcommand{\pos}[1]{[#1]_{+}}
\newcommand{\gmin}{\gamma_{\min}}
\newcommand{\gmax}{\gamma_{\max}}
\newcommand{\Dg}{\Delta\gamma}

\newcommand{\thmref}[1]{Theorem~\ref{#1}}
\newcommand{\lemref}[1]{Lemma~\ref{#1}}

\newcommand{\obsref}[1]{Observation~\ref{#1}}
\newcommand{\appref}[1]{Appendix~\ref{#1}}
\newcommand{\secref}[1]{Section~\ref{#1}}
\newcommand{\RightAlign}{\emph{Right-Align}}
\newcommand{\SignedUpdate}{\emph{Signed-Update}}

\title{\textbf{Cooling Noisy Population Protocols}\\
Vanishing-Gain Averaging and Localisation under Persistent Errors}

\author[1]{Leszek G\k asieniec}

\author[2]{\L ukasz Kuszner}

\author[1]{Paul Spirakis}

\author[3]{Grzegorz Stachowiak}

\affil[1]{University of Liverpool, United Kingdom}

\affil[2]{University of Gdańsk, Poland}

\affil[3]{University of Wrocław, Poland}

\date{}

\begin{document}
\pagenumbering{gobble}


\maketitle 

\vspace{14mm}
\begin{abstract}
We study noise-tolerant coordination in anonymous population protocols whose agents interact in uniformly random pairs, maintain real-valued labels, and have no global clock. Each agent cools its updates using a gain $\gamma_k=(k+1)^{-p}$ computed from its own local interaction counter. First, for pairwise averaging with independent communication errors drawn from a fixed centred finite-variance distribution, we prove convergence with probability 1 to a common finite random value for every $1/2<p\le1$, and derive convergence-rate exponents under the same assumptions. Second, for one-dimensional localisation with fixed non-degenerate Gaussian displacement errors, we show that the one-sided \RightAlign{} rule rectifies centred noise and cannot stabilise; a signed update based on the relative-position discrepancy restores convergence to the true agents locations up to one finite random translation. Third, we analyse the setting in which both communication errors and displacement-measurement errors depend on the physical distance between the interacting devices. For additive independent Gaussian errors, the two sources combine into one centred effective error whose variance is the sum of the distance-dependent component variances. The signed updates protocol remains convergent for every fixed finite geometry, and the average combined variance determines the leading common-translation fluctuation. These results distinguish cooling-compatible signed corrections from one-sided updates that turn persistent noise into drift.
\end{abstract}

\vfill



\pagenumbering{arabic}

\section{Introduction}
\label{sec:introduction}

Population protocols are a foundational model for distributed computation by large collections of anonymous, resource-limited agents that interact in pairs \cite{AngluinEtAl2006}. In the classical model, an interaction applies a fixed transition function to the states of an ordered pair: the \emph{initiator} and the \emph{responder}, and the finite state space is independent of the population size. A protocol starts from an input configuration and must stabilise to configurations encoding a valid output. The computational power of this model and its basic  properties are well understood \cite{AngluinAER2007,AspnesRuppert2009}.

We use the standard uniform random scheduler: at each step, an ordered pair of distinct agents is chosen uniformly and independently of the past. The \emph{sequential time} is the number of interactions, while the \emph{parallel time} is the sequential time divided by the population size $n$. State complexity and parallel stabilisation time are the two customary efficiency measures in probabilistic population protocols; epidemic spreading, concentration, and martingale arguments are central tools in their analysis \cite{AngluinAspnesEisenstat2008,BerenbrinkEtAl2018}. Our protocols retain anonymity, random pairwise interactions, and the absence of a common clock, but allow real-valued labels and unbounded local counters, as required by averaging and spatial localisation. Each gain is computed from the updating agent's own counter: agents need neither identifiers nor knowledge of $n$ or of global time.

\subsection{Noisy averaging and local cooling}

Pairwise averaging is a basic continuous-state coordination task and a canonical example of randomised gossip \cite{BoydEtAl2006,OlfatiSaberFaxMurray2007}. In graph-based consensus models, persistent communication noise creates a familiar tension: continued updates reduce disagreement but also inject fresh variance. Decreasing stepsizes are a classical remedy, and their convergence and rate trade-offs have been analysed for noisy measurements, random link failures, and additive channel noise \cite{HuangManton2009,KarMoura2009,MosqueraEtAl2010}.

Here interactions are generated by a random scheduler, agents have no shared round counter, and simultaneous gains are therefore slightly different. Previous work on noisy population-protocol averaging shows that the population may remain tightly clustered around its current mean while the mean itself performs an unbounded noise-driven walk \cite{MallmannTrennMausPajak2019}. Section~\ref{sec:averaging} asks whether stabilisation can be recovered using only local information. We show that the local gain
\[
  \gamma_i(t)=(c_i(t)+1)^{-p},\qquad \frac12<p\le1,
\]
where $c_i(t)$ is the participation counter of agent $i$, is sufficient. Our contribution is an implementation and analysis of local-counter cooling under fresh persistent errors, with separate rates for disagreement and common-value motion.

\subsection{Spatial localisation and noisy alignment}

Spatial population protocols enrich an interaction with geometric information, such as a distance or a relative displacement vector \cite{GasieniecEtAl2025}. In one dimension, exact localisation can be obtained from the update
\begin{equation}
 x_i\leftarrow\max\{x_i,x_r-v_{ir}\},\qquad v_{ir}=p_r-p_i,
 \label{eq:intro-max}
\end{equation}
where $p_i$ is the fixed physical position of agent $i$. We call this operation \RightAlign{}. With translation errors $y_i=x_i-p_i$, the update becomes $y_i\leftarrow\max\{y_i,y_r\}$; the maximum translation spreads epidemically and yields exact stabilisation in $O(\log n)$ parallel time with high probability \cite{GasieniecEtAl2025}. This is a real-valued self-stabilising scenario~\cite{Dijkstra1974,Dolev2000,AngluinAspnesFischerJiang2008} in which arbitrary initial labels are driven to a legitimate coordinate system up to one global translation.

Persistent errors fundamentally alter the computation. At consistency, the one-sided \RightAlign{} rule accepts positive perturbations but discards negative ones, thereby converting centred noise into positive drift. Section~\ref{sec:localisation} first isolates this effect under a fixed error distribution. We observe that noisy \RightAlign{} does not converge and that power-law damping alone cannot simultaneously suppress the rectification bias and preserve enough total correction. Replacing the one-sided correction by the signed relative-position discrepancy yields a convergent algorithm.

Distributed localisation and consensus under noisy relative measurements have also been studied in graph-based sensor and multi-agent models \cite{DiaoEtAl2013,FranchiOrioloStegagno2013,WangEtAl2023}. 
A related asynchronous gossip method uses local counters to
time-average a constant-gain process without a shared clock,
obtaining the least-squares solution for a fixed noisy measurement
vector\ \cite{RavazziEtAl2013}. In our model, errors are resampled at each encounter, cooling makes the labels themselves converge, and the analysis separates relative and common-translation rates. Other work allows noise intensity to depend on relative state or distance \cite{LiWuZhang2014,QiuLiuXie2016}. Here interactions are anonymous and uniformly random; the relevant distance is the fixed physical separation of the devices, so noise need not disappear at geometric consistency.

\subsection{From fixed error laws to joint distance-dependent noise}

Sections~\ref{sec:averaging} and~\ref{sec:localisation} use fixed, pair-independent error distributions. This cleanly separates the two computational mechanisms: cooling an averaging process and replacing noisy one-sided alignment by a signed update. Section~\ref{sec:physical-scenarios} then treats the most realistic model considered here. The responder's transmitted label and the measured relative displacement are both corrupted, and the law of each error source depends on the physical distance of the interacting pair. For additive errors, the algorithm observes only their difference. Under independent Gaussian errors, the variances add, so both communication quality and sensing quality enter the same effective error. This reduction permits a single convergence analysis while keeping the two physical sources explicit in the geometry-dependent constants.

The distance-dependent model is not intended as a universal hardware law. The communication and sensing variance profiles are inputs to the model. Convergence with probability 1 requires conditional centring and bounded second moments; The assumption of Gaussian sources of errors is used only for sharper rate constants and the explicit \RightAlign{} drift calculation.

\subsection{Contributions}

The paper has three main contributions.
All protocols are uniform and population-size-oblivious: their transition rules do not depend on $n$, which is used only in the analysis of the protocols. 
\begin{enumerate}
\item \textbf{Noisy averaging with local cooling (Section~\ref{sec:averaging}).}
For independent communication errors drawn from a fixed centred distribution with finite variance, the locally cooled averaging protocol converges with probability 1 to a common finite random value for every fixed population and every $1/2<p\le1$ (Theorem~\ref{thm:avg}). Under the same assumptions, the mean-square disagreement is $O(k^{-p})$ and the 
RMS distance of the population mean from its limiting value
has order $k^{1/2-p}$ when the noise variance is positive (Theorem~\ref{thm:avg-rate}), where $k$ is the local-count time scale.
This complements~\cite{MallmannTrennMausPajak2019} by replacing the drifting common mean with
convergence with probability 1 using gains computed solely from local interaction counters.

\item \textbf{Stabilising noisy one-dimensional localisation (Section~\ref{sec:localisation}).}
For fixed non-degenerate Gaussian displacement errors, plain \RightAlign{} has positive expected drift and arbitrarily large relative-coordinate excursions (Observation~\ref{obs:maxfail}). Cooling the one-sided increment cannot recover both noise robustness and exact-case convergence. A signed update based on the relative-position discrepancy converges with probability 1 to $x_i=p_i+C_\infty$ for a finite random translation $C_\infty$ and has the same two-mode rate exponents as averaging (Theorems~\ref{thm:signed} and~\ref{thm:signed-rate}).
This extends the exact-displacement setting of~\cite{GasieniecEtAl2025} to persistent
measurement errors and identifies a signed update that restores
asymptotic localisation.

\item \textbf{Joint distance-dependent communication and measurement errors (Section~\ref{sec:physical-scenarios}).}
We allow both the communicated label and the measured displacement to be corrupted by distance-dependent errors. For independent Gaussian sources with variances $s_c^2(d)$ and $s_m^2(d)$, the effective variance is $q(d)=s_c^2(d)+s_m^2(d)$. The signed protocol converges for every fixed finite geometry. The worst-pair variance controls the general bounds, while the average combined variance $\bar q$ determines the leading common-translation fluctuation (Theorems~\ref{thm:effective-signed}--\ref{thm:joint-rate}). The one-sided rule remains non-convergent whenever the effective variances are bounded away from zero (Observation~\ref{obs:joint-max}).
\end{enumerate}

\subsection{Organisation}

Section~\ref{sec:model} introduces the model and common probabilistic tools. Sections~\ref{sec:averaging} and~\ref{sec:localisation} establish the fixed-distribution results. Section~\ref{sec:physical-scenarios} develops the joint distance-dependent model, and Section~\ref{sec:conclusion} concludes. The appendices contain all detailed proofs omitted from the main text, including the complete Right-Align impossibility arguments, convergence derivations, and rate analyses.

\section{Model and probabilistic preliminaries}
\label{sec:model}

We consider a population of $n\ge2$ agents, arbitrary finite deterministic initial labels, and local counters initialised to zero. Agent indices are used only for analysis. 

Sequential time $t=0,1,2,\ldots$ is the number of completed interactions, so the update at interaction $t$ takes the state from $x(t)$ to $x(t+1)$. Parallel time is $t/n$. Let $\F_t$ be the sigma-field generated by the initial configuration and all pair choices and errors in interactions $0,\ldots,t-1$. Thus $x(t)$ and all counters $c_i(t)$ are $\F_t$-measurable. Pair selections are independent and uniform. Sections~\ref{sec:averaging} and~\ref{sec:localisation} use independent fresh errors from fixed, pair-independent distributions. In \secref{sec:physical-scenarios}, both communication and measurement error laws depend on the selected pair through its fixed physical separation; the scheduler remains independent and uniform. That section also states the conditional moment assumptions needed beyond independent Gaussian errors.

\subsection{Analytical framework}

Let $x_i(t)$ be the label of agent $i$ after $t$ completed interactions, immediately before interaction $t$.
A sample path \emph{converges} if for every $\varepsilon>0$ there is $t'$ such that for every agent $i$ and every interaction $t>t'$
\begin{equation}
    |x_i(t)-x_i (t')|<\varepsilon.
\end{equation}

For a noisy protocol, \emph{stabilisation} means convergence with probability 1 of the labels to a finite configuration satisfying the task specification. Label updates may persist at every finite time.

Our analysis uses standard stochastic-approximation tools adapted to the population-protocol setting \cite{RobbinsMonro1951,RobbinsSiegmund1971,Benaim1999,KushnerYin2003}. The main additional difficulty is that agents have no common iteration counter: each maintains its own local counter, so gains are random and slightly different at the same global time. Concentration and Borel--Cantelli arguments control this mismatch, after which Lyapunov and martingale methods establish convergence.

\subsection{Schedulers, local counters, and gains}
\label{subsec:counters-main}

For averaging, both agents update and the interaction orientation is irrelevant. We therefore describe the scheduler as choosing an unordered pair uniformly from the $\binom n2$ possibilities. The counter $c_i(t)$ records the number of participations by $i$ before interaction $t$.

For localisation, only the initiator updates. Write $(I_t,R_t)$ for the ordered pair selected uniformly from the $n(n-1)$ distinct initiator--responder pairs. Here $c_i(t)$ records the number of initiations by $i$ before interaction $t$. In either protocol the updating agent uses its pre-interaction counter to compute
\begin{equation}
 \gamma_i(t)=(c_i(t)+1)^{-p},\qquad \frac12<p\le1,
 \label{eq:gain}
\end{equation}
and increments its counter after the update. For comparison across agents, define
\[
 \gmin(t)=\min_i\gamma_i(t),\qquad
 \gmax(t)=\max_i\gamma_i(t),\qquad
 \Dg(t)=\max_{i,j}|\gamma_i(t)-\gamma_j(t)|.
\]
The probability that a given agent's counter increases at one interaction is
\begin{equation}
 \lambda_{\rm av}=\frac2n\quad\text{for averaging},\qquad
 \lambda_{\rm loc}=\frac1n\quad\text{for localisation}.
 \label{eq:lambdas}
\end{equation}

\begin{lemma}[Counter concentration]
\label{lem:counter-concentration}
Fix either scheduler and write $\lambda$ for the corresponding value in \eqref{eq:lambdas}.
For every $\delta>0$, with probability 1 there is a finite random $T$ such that,
simultaneously for all agents and all $t\ge T$,
\begin{equation}
  |c_i(t)-\lambda t|\le t^{1/2+\delta}.
  \label{eq:counterconc}
\end{equation}
Moreover, for every fixed $t\ge1$,
\begin{equation}
 \Pp\!\left(\max_i|c_i(t)-\lambda t|>t^{1/2+\delta}\right)
 \le 2n\exp(-2t^{2\delta}).
 \label{eq:counter-tail}
\end{equation}
\end{lemma}
\noindent The proof is given in \appref{app:counter-gain-proofs}.

\begin{lemma}[Gain comparability and summable mismatch]
\label{lem:gain-control}
Let $1/2<p\le1$ and fix $0<\delta<\min\{1/2,p-1/2\}$. With probability 1, for all sufficiently
large $t$,
\begin{align}
  \gamma_i(t)&=(\lambda t)^{-p}\bigl(1+O(t^{-1/2+\delta})\bigr)
       \quad\text{uniformly in }i, \label{eq:gain-asymptotic}\\
  \Dg(t)&=O(t^{-p-1/2+\delta}). \label{eq:mismatch}
\end{align}
Consequently,
\begin{equation}
 \sum_{t\ge0}\gmin(t)=\infty,
 \qquad
 \sum_{t\ge0}\gmax(t)^2<\infty,
 \qquad
 \sum_{t\ge0}\Dg(t)<\infty
 \quad\text{with probability 1.}
 \label{eq:gainseries}
\end{equation}
\end{lemma}
\noindent The proof is given in \appref{app:counter-gain-proofs}.

The summability of the gain mismatch is the point that allows purely local counters to replace a common global clock.

\subsection{Stochastic tools}
\label{subsec:tools-main}

The standard stochastic-approximation conditions are
$\sum_k\gamma_k=\infty$ and $\sum_k\gamma_k^2<\infty$.
They balance persistent correction against finite accumulated quadratic variation
\cite{RobbinsMonro1951,Benaim1999,KushnerYin2003}. We use the following
almost-supermartingale theorem \cite{RobbinsSiegmund1971}.

\begin{lemma}[Robbins--Siegmund]
\label{lem:RS}
Let $Z_t,A_t,B_t,C_t$ be nonnegative adapted random variables satisfying
\[
 \E[Z_{t+1}\mid\F_t]\le(1+A_t)Z_t-B_t+C_t.
\]
If $\sum_tA_t<\infty$ and $\sum_tC_t<\infty$ with probability 1, then $Z_t$ converges with probability 1 to a finite limit and $\sum_tB_t<\infty$ with probability 1.
\end{lemma}

We also use the following elementary martingale criterion.

\begin{lemma}[Square-summable martingale series]
\label{lem:mart-series}
Let $D_t$ be $\F_{t+1}$-measurable martingale differences with $\E[D_t\mid\F_t]=0$ and finite conditional second moments. If
\[
 \sum_t\E[D_t^2\mid\F_t]<\infty\qquad\text{with probability 1},
\]
then $\sum_tD_t$ converges with probability 1. If the sum of the unconditional second moments is
finite, then the convergence also holds in $L^2$ and
\[
 \left\|\sum_{s\ge t}D_s\right\|_2^2=\sum_{s\ge t}\E[D_s^2].
\]
\end{lemma}

For the mean-square rates we repeatedly reduce to a deterministic scalar recursion.

\begin{lemma}[Rate comparison]
\label{lem:rate-comparison}
Suppose $u_t\ge0$ and, for some $a,b>0$,
\begin{equation}
 u_{t+1}\le\bigl(1-a t^{-p}+o(t^{-p})\bigr)u_t+b t^{-2p}+r_t,
 \label{eq:rate-rec-general}
\end{equation}
where $r_t$ decreases faster than every inverse polynomial. If $1/2<p<1$, then
$u_t=O(t^{-p})$. If $p=1$ and $a>1$, then $u_t=O(t^{-1})$.
\end{lemma}

Finally, the tails of the squared gains satisfy
\begin{equation}
 \sum_{m=k}^{\infty}(m+1)^{-2p}
 =\frac{k^{1-2p}}{2p-1}(1+o(1)).
 \label{eq:ptail}
\end{equation}
Proofs of Lemmas~\ref{lem:mart-series} and~\ref{lem:rate-comparison} appear in \appref{app:prob-tools}, together with a weighted-event lemma for the damping analysis and moment bounds for controlling rare counter-imbalance events in the rate proofs.

\section{Noisy averaging with local cooling}
\label{sec:averaging}

This section establishes our first contribution: noisy averaging with local cooling, including its convergence rates. Communication errors are sampled from one fixed distribution, independently of the selected pair. The proof separates disagreement from the population mean, because convergence of the former alone does not imply stabilisation.

\subsection{Protocol and convergence}
At interaction $t$, let $\{i,j\}$ be the selected pair. We use the sequential two-message noise model studied for population-protocol averaging in \cite{MallmannTrennMausPajak2019}: the two directed messages are perturbed independently. Thus $\xi_i(t),\xi_j(t)$ are independent samples from a fixed distribution $\nu_{\rm av}$ with mean zero and variance $\sigma^2<\infty$. Samples are independent across interactions and of the entire pair schedule. With
$a=\gamma_i(t)$ and $b=\gamma_j(t)$, we update
\begin{align}
 x_i^+&=x_i+\frac a2\bigl(x_j-x_i+\xi_i(t)\bigr),
 \label{eq:avg-i}\\
 x_j^+&=x_j+\frac b2\bigl(x_i-x_j+\xi_j(t)\bigr).
 \label{eq:avg-j}
\end{align}
Both right-hand sides use the pre-interaction values; all other labels remain unchanged. For $a=b=1$ this is the usual noisy pairwise averaging rule. Define the population mean,
centred values, and disagreement by
\begin{equation}
 M(t)=\frac1n\sum_{k=1}^nx_k(t),\qquad
 z_k(t)=x_k(t)-M(t),\qquad
 V(t)=\sum_{k=1}^nz_k(t)^2.
 \label{eq:avgMV}
\end{equation}

\begin{theorem}[Stabilisation of noisy averaging with probability 1]
\label{thm:avg}
Assume independent uniform unordered-pair interactions and communication errors sampled independently from the fixed centred finite-variance distribution $\nu_{\rm av}$. For all $n\ge2$ and every
$1/2<p\le1$, the locally cooled protocol \eqref{eq:avg-i}--\eqref{eq:avg-j} admits a finite
random variable $X_\infty$ such that
\begin{equation}
 x_i(t)\longrightarrow X_\infty\qquad\text{for every }i
 \quad\text{with probability 1.}
 \label{eq:avg-limit}
\end{equation}
The limit need not equal the initial population average.
\end{theorem}


\begin{lemma}[Averaging one-step estimate]
\label{lem:avg-one-step}
Under \eqref{eq:avg-i}--\eqref{eq:avg-j}, the disagreement satisfies
\[
 \E[V(t+1)\mid\F_t]
 \le\left(1+\frac{\Dg(t)}2+\frac{\gmax(t)^2}{n-1}\right)V(t)
 -\frac{2\gmin(t)}{n-1}V(t)+\frac{\sigma^2}{2}\gmax(t)^2.
\]
\end{lemma}

The complete proof of \lemref{lem:avg-one-step} and \thmref{thm:avg} is in
\appref{app:avg-convergence-proof}. The estimate separates the contracting
pairwise term from the summable local-gain mismatch and the square-summable
noise term. Robbins--Siegmund gives $V(t)\to0$ with probability 1. The population
mean is then decomposed into an absolutely summable gain-asymmetry term and
a martingale series with summable conditional variance.

\subsection{Rates}
\label{subsec:avg-rate}
The rates require only the finite-variance assumptions used for convergence.
Conditioning on the pair schedule controls rare counter-imbalance events through a second-moment bound.

\begin{theorem}[Averaging rate]
\label{thm:avg-rate}
Under the assumptions of \thmref{thm:avg},
\begin{equation}
 \E V(t)=O(t^{-p}).
 \label{eq:avg-rate-global}
\end{equation}
Consequently, for every pair $i,j$,
\begin{equation}
 \bigl(\E[(x_i(t)-x_j(t))^2]\bigr)^{1/2}=O(t^{-p/2}).
 \label{eq:avg-relrate}
\end{equation}
Writing $k=\lambda_{\rm av}t=2t/n$ for the deterministic local-count time scale gives the exponents $k^{-p}$ and $k^{-p/2}$, respectively. With probability 1, every actual local counter is asymptotic to this scale. This deterministic rescaling is distinct from the random completion time used below.

Let
\[
 \tau_k^{\rm av}=\inf\{t:\min_i c_i(t)\ge k\}.
\]
Write $M(\infty)=X_\infty$. Then the full 
RMS distance of the population mean from its limiting value satisfies, for every fixed
$0<\delta<\min\{p-1/2,p/2\}$,
\begin{equation}
 \|M(\infty)-M(\tau_k^{\rm av})\|_2
 \le
 \frac{\sigma}{2\sqrt{n(2p-1)}}k^{1/2-p}(1+o(1))
 +O\!\left(k^{1/2-3p/2+\delta}\right).
 \label{eq:avg-commonrate}
\end{equation}
When $\sigma>0$, the noise-martingale tail has root mean square (RMS) asymptotic to the first term, and the gain-mismatch correction is of strictly smaller order. Thus the full 
RMS distance of the population mean from its limiting value has
order $k^{1/2-p}$.
\end{theorem}

The complete proof of \thmref{thm:avg-rate}, including the noise-tail asymptotic and the rare-event truncation that justifies
passing from random local counters to the deterministic scalar recursion of
\lemref{lem:rate-comparison}, is in \appref{app:avg-rate-proof}.

The two rates describe different aspects of stabilisation. Relative
disagreement measures how closely the agents agree \emph{now}, whereas
the RMS distance of the population mean from its limiting value measures
the size of the remaining net change in their average.
For $1/2<p<1$, the latter decays more slowly. At $p=1$, both displayed RMS exponents equal $-1/2$; a numerical exponent summary appears in Table~\ref{tab:p} in the appendix. This is an asymptotic comparison within the specified power-law family, not a claim of optimal finite-time performance over all possible gain schedules.

\section{Stabilising noisy one-dimensional localisation}
\label{sec:localisation}

We now establish the second contribution, using exact communication and displacement errors drawn from a fixed distribution independent of the interacting pair and its physical separation. This isolates the effect of the update rule before the joint distance-dependent model of \secref{sec:physical-scenarios}. Let $p_1,\ldots,p_n\in\R$ be fixed distinct physical positions. A label vector is geometrically
consistent exactly when there exists one translation $C$ such that $x_i=p_i+C$ for all $i$.
Define translation errors
\begin{equation}
 y_i=x_i-p_i.
 \label{eq:y}
\end{equation}
Thus localisation is equivalent to consensus among the $y_i$.

\subsection{Exact-displacement baseline}
With the exact relative displacement $v_{ir}=p_r-p_i$, the update studied in \cite{GasieniecEtAl2025},
\[
 x_i^+=\max\{x_i,x_r-v_{ir}\},
\]
becomes
\begin{equation}
 y_i^+=\max\{y_i,y_r\}.
 \label{eq:max-exact}
\end{equation}
The current maximum translation is invariant and spreads epidemically. This noiseless baseline is analysed in \cite{GasieniecEtAl2025}. Our focus is the effect of persistent errors in the information used to form the relative-position discrepancy.

\subsection{Fixed-variance displacement noise}
\label{subsec:fixed-noise}
Assume
\begin{equation}
 \widehat v_{ir}=v_{ir}+\xi_t,
 \qquad \xi_t\sim N(0,\sigma^2),\quad \sigma>0,
 \label{eq:loc-fixed}
\end{equation}
independently at each interaction and independently of the selected pair, and put $\eta_t=-\xi_t$. The same distribution $N(0,\sigma^2)$ is used for all pairs, irrespective of their physical separation. Plain \RightAlign{} becomes
\begin{equation}
 y_i^+=\max\{y_i,y_r+\eta_t\}.
 \label{eq:max-noisy}
\end{equation}
Let
\[
 H(t)=\max_i y_i(t),\qquad R(t)=\max_i y_i(t)-\min_i y_i(t).
\]

\begin{observation}[Plain noisy \RightAlign{} cannot stabilise]
\label{obs:maxfail}
Under \eqref{eq:max-noisy},
\begin{equation}
 \E[H(t+1)-H(t)\mid\F_t]\ge\frac{\sigma}{n\sqrt{2\pi}}.
 \label{eq:maxdrift}
\end{equation}
Consequently $\E H(t)\ge H(0)+t\sigma/(n\sqrt{2\pi})$. Moreover,
\begin{equation}
 H(t)\longrightarrow+\infty\quad\text{with probability 1},
 \qquad
 \limsup_{t\to\infty}R(t)=\infty\quad\text{with probability 1}
 \label{eq:max-as-fail}
\end{equation}
Consequently, the labels do not converge to a finite configuration, and no band $R(t)\le A$ with fixed $A<\infty$ is eventually invariant.
\end{observation}

The proof of \obsref{obs:maxfail} is in \appref{app:right-align-plain}. Its key observation is that, when the responder is a current maximiser, a positive error increases the maximum by at least that error. At exact consistency, the one-sided increment is $[\eta_t]_+$, whose expectation is $\sigma/\sqrt{2\pi}>0$. Thus centred measurement noise becomes a systematic upward correction.

\subsection{Why cooling \RightAlign{} is insufficient}
Consider the damped one-sided rule
\begin{equation}
 y_i^+=y_i+\gamma_i(t)[y_r-y_i+\eta_t]_+.
 \label{eq:dampedmax}
\end{equation}

\begin{observation}[Nonsummable damping does not remove \RightAlign{} drift]
\label{obs:damped-max}
For $1/2<p\le1$ under fixed non-degenerate Gaussian noise, the event
\[
 \{\exists C\in\R:\ y_i(t)\to C\text{ for every }i\}
\]
has probability zero.
\end{observation}

\begin{observation}[Summable gains lose the exact-case guarantee]
\label{obs:summable-max-counterexample}
For $p>1$ there is a three-agent noiseless initial configuration for which
the damped \RightAlign{} protocol fails to converge to a common translation
with positive probability under the uniform random scheduler.
\end{observation}

The proofs, including the explicit counterexample, are in
\appref{app:right-align-damping}. Together the observations identify the
power-law obstruction: nonsummable gains accumulate rectification, whereas
summable gains can leave a permanent noiseless discrepancy.

\subsection{\SignedUpdate{}: protocol and convergence}
\label{subsec:signed-protocol}
We replace \RightAlign{} by \SignedUpdate{}, which retains the sign of the relative-position discrepancy. When initiator $i$ interacts with responder $r$, define
\begin{equation}
 \delta_{ir}(t):=x_r(t)-\widehat v_{ir}(t)-x_i(t).
 \label{eq:fixed-relative-position-discrepancy}
\end{equation}
The quantity $x_r-\widehat v_{ir}$ is the coordinate for $i$ implied by the information received from $r$, and $\delta_{ir}$ measures its discrepancy from $i$'s current coordinate. We call $\delta_{ir}$ the \emph{relative-position discrepancy}. The signed update is
\begin{equation}
 x_i^+=x_i+\gamma_i(t)\delta_{ir}(t)
      =x_i+\gamma_i(t)\bigl(x_r-\widehat v_{ir}-x_i\bigr).
 \label{eq:signedx}
\end{equation}
Only the initiator updates. In translation coordinates,
\begin{equation}
 \delta_{ir}(t)=y_r(t)-y_i(t)-\xi_t,
 \qquad
 y_i^+=y_i+\gamma_i(t)\delta_{ir}(t).
 \label{eq:signedy}
\end{equation}
Thus the discrepancy consists of the true alignment error $y_r-y_i$ and the effective measurement error $-\xi_t$. At consistency, the stochastic increment is $-\gamma_i(t)\xi_t$, with zero conditional mean and variance $\gamma_i(t)^2\sigma^2$. Unlike \RightAlign{}, this rule preserves the centring of the measurement error.

\begin{theorem}[Signed localisation with probability 1]
\label{thm:signed}
Under independent uniform ordered interactions, the fixed-variance model
\eqref{eq:loc-fixed}, and local initiator gains \eqref{eq:gain} with $1/2<p\le1$, there is a
finite random variable $C_\infty$ such that
\begin{equation}
 x_i(t)\longrightarrow p_i+C_\infty
 \qquad\text{for every }i
 \quad\text{with probability 1.}
 \label{eq:loclimit}
\end{equation}
\end{theorem}

Let
\[
 M(t)=\frac1n\sum_i y_i(t),\qquad z_i(t)=y_i(t)-M(t),\qquad
 V(t)=\sum_i z_i(t)^2.
\]


\begin{lemma}[Signed-localisation one-step estimate]
\label{lem:loc-one-step}
Under \eqref{eq:signedy}, the disagreement satisfies
\[
 \E[V(t+1)\mid\F_t]
 \le\left(1+\frac{2\gmax(t)^2}{n}\right)V(t)
 -\frac{2\gmin(t)}{n-1}V(t)+\sigma^2\gmax(t)^2.
\]
\end{lemma}

The complete proof of \lemref{lem:loc-one-step} and \thmref{thm:signed}
is in \appref{app:signed-convergence-proof}. Robbins--Siegmund yields
$V(t)\to0$. The translation mean is the sum of a square-summable noise
martingale, a pair-selection martingale, and an absolutely summable
predictable term caused by the mismatch between local gains.

\subsection{Convergence rates}
\label{subsec:loc-rate}

The signed protocol has the same two-mode rate structure as noisy averaging. Here $k=\lambda_{\rm loc}t=t/n$ is the deterministic local-count time scale, and the random time below marks the completion of at least $k$ initiations by every agent.

\begin{theorem}[Signed-localisation rate]
\label{thm:signed-rate}
Assume the hypotheses of \thmref{thm:signed}. Then
\begin{equation}
 \E V(t)=O(t^{-p}),
 \qquad
 \bigl\|(x_i(t)-x_j(t))-(p_i-p_j)\bigr\|_2=O(t^{-p/2}).
 \label{eq:locrelrate}
\end{equation}
Let
\[
 \tau_k^{\rm loc}=\inf\{t:\min_i c_i(t)\ge k\}.
\]
For every fixed $0<\delta<\min\{p-1/2,p/2\}$,
\begin{equation}
 \|C_\infty-M(\tau_k^{\rm loc})\|_2
 \le
 \frac{\sigma}{\sqrt{n(2p-1)}}k^{1/2-p}(1+o(1))
 +O\!\left(k^{1/2-3p/2+\delta}\right).
 \label{eq:loccommonrate}
\end{equation}
The noise-martingale tail has RMS asymptotic to the first term, so the common-translation tail has RMS order $k^{1/2-p}$. It is the slower mode for $1/2<p<1$, while both displayed RMS exponents equal $-1/2$ at $p=1$.
\end{theorem}

A full proof, including the $L^2$ estimates for the pair-selection and gain-mismatch tails, is
given in \appref{app:signed-rate-proof}.

\section{Localisation with distance-dependent combined errors}
\label{sec:physical-scenarios}

Our third contribution treats simultaneous distance-dependent communication
and sensing errors. Both the communicated responder label and the measured
relative displacement are corrupted, with error laws determined by the
physical separation of the interacting agents. The additive-noise model
and its scope are specified below.

\subsection{Joint distance-dependent noise model}
\label{subsec:both-noisy}
\label{subsec:joint-distance-model}

Keep the fixed one-dimensional geometry of \secref{sec:localisation} and put
\[
 d_{ir}=|p_i-p_r|.
\]
For distinct positions in a fixed finite population,
\begin{equation}
 0<d_{\min}:=\min_{i\ne r}d_{ir}
 \le d_{ir}\le d_{\max}:=\max_{i\ne r}d_{ir}<\infty.
 \label{eq:dminmax}
\end{equation}
In an ordered interaction $(i,r)$, the initiator receives
\begin{equation}
 \widetilde x_r=x_r+\zeta_t
 \label{eq:comm-noise}
\end{equation}
and measures
\begin{equation}
 \widetilde v_{ir}=v_{ir}+\varepsilon_t,
 \qquad v_{ir}=p_r-p_i.
 \label{eq:sense-noise}
\end{equation}
The measurement is still a \emph{signed relative displacement}, not an
unsigned range; the distance $d_{ir}$ determines the error laws. In the
basic joint model,
\begin{equation}
 \begin{aligned}
  \zeta_t\mid(\F_t,I_t=i,R_t=r)&\sim N\!\left(0,s_c^2(d_{ir})\right),\\
  \varepsilon_t\mid(\F_t,I_t=i,R_t=r)&\sim N\!\left(0,s_m^2(d_{ir})\right).
 \end{aligned}
 \label{eq:scenario-both}
\end{equation}
The two errors are independent conditional on the selected pair, and fresh
errors are sampled independently across interactions conditional on the
pair schedule. The functions $s_c^2(d)$ and $s_m^2(d)$ specify communication
and measurement variances, respectively. They may have different forms and need only be nonnegative and
finite at the separations present in the population. Nondecreasing
functions model increasing uncertainty with distance, but monotonicity is
not needed for the proofs. Constant functions or an identically zero
source give simpler submodels.

The same relative-position discrepancy now satisfies
\begin{equation}
 \delta_{ir}(t)
 :=\widetilde x_r(t)-\widetilde v_{ir}(t)-x_i(t)
 =y_r(t)-y_i(t)+\eta_t,
 \qquad \eta_t:=\zeta_t-\varepsilon_t.
 \label{eq:effective-noise}
\end{equation}
Here $\eta_t$ is the \emph{effective error}: it is the random component of the discrepancy, not the discrepancy itself.
Independence therefore gives
\begin{equation}
 q_{ir}=s_c^2(d_{ir})+s_m^2(d_{ir}),
 \label{eq:variance-adds}
\end{equation}
with
\begin{equation}
 \eta_t\mid(\F_t,I_t=i,R_t=r)\sim N(0,q_{ir}).
 \label{eq:effective-gaussian}
\end{equation}

\begin{lemma}[Effective-error reduction]
\label{lem:effective-error}
For every ordered interaction $(i,r)$, both \RightAlign{} and \SignedUpdate{} depend on the two additive error sources only through
\[
  \eta_t=\zeta_t-\varepsilon_t.
\]
Under the conditionally independent Gaussian model \eqref{eq:scenario-both}, the effective error is conditionally centred Gaussian with variance $q_{ir}$ given by \eqref{eq:variance-adds}. More generally, if the two component errors are conditionally centred and have finite conditional second moments, then $\eta_t$ is conditionally centred and
\[
 \E[\eta_t^2\mid\F_t,I_t=i,R_t=r]
 =\Var(\zeta_t-\varepsilon_t\mid\F_t,I_t=i,R_t=r).
\]
\end{lemma}

\begin{proof}
Substituting \eqref{eq:comm-noise} and \eqref{eq:sense-noise} gives
\[
 \widetilde x_r-\widetilde v_{ir}-x_i
 =(x_r-v_{ir}-x_i)+(\zeta_t-\varepsilon_t)
 =y_r-y_i+\eta_t.
\]
Thus the protocol sees only the difference of the two errors. Conditional centring follows by linearity of conditional expectation. Under conditional independence, variances add; the difference of independent centred Gaussians is Gaussian, proving \eqref{eq:variance-adds}--\eqref{eq:effective-gaussian}.
\end{proof}
Define the geometry-dependent constants
\begin{equation}
 \begin{aligned}
  q_{\min}&=\min_{i\ne r}q_{ir},\qquad
  q_{\max}=\max_{i\ne r}q_{ir}<\infty,\\
  \bar q&=\frac{1}{n(n-1)}\sum_{i\ne r}q_{ir}.
 \end{aligned}
 \label{eq:joint-variance-constants}
\end{equation}
For a fixed geometry, the conditional law for each pair is fixed in time,
but different pair distances may give different laws. This contrasts with
the same pair-independent distribution used at every interaction in the
preceding sections. The noise depends on physical separation, not on
computed-label disagreement, and thus need not disappear at consistency.
Neither the true distance nor the variance functions are needed to compute
the local cooling gain.

\subsection{Stabilisation under combined distance-dependent errors}
\label{subsec:joint-convergence}

With both information sources noisy, \SignedUpdate{} becomes
\begin{equation}
 x_i^+=x_i+\gamma_i(t)\delta_{ir}(t)
      =x_i+\gamma_i(t)
       \bigl(\widetilde x_r-\widetilde v_{ir}-x_i\bigr),
 \label{eq:joint-signed-x}
\end{equation}
or, equivalently,
\begin{equation}
 y_i^+=y_i+\gamma_i(t)\delta_{ir}(t)
      =y_i+\gamma_i(t)(y_r-y_i+\eta_t).
 \label{eq:effective-signed}
\end{equation}
Only the initiator updates. Retain the translation mean $M$, centred
translation errors $z_i$, and disagreement $V$ from
\secref{subsec:signed-protocol}.

\begin{lemma}[Joint signed one-step estimate]
\label{lem:joint-one-step}
Assume that the effective error is conditionally centred and satisfies
\[
 \E[\eta_t^2\mid\F_t,I_t=i,R_t=r]\le q_{\max}
\]
for every pair and time. Then the update \eqref{eq:effective-signed} satisfies
\begin{equation}
 \E[V(t+1)\mid\F_t]
 \le\left(1+\frac{2\gmax(t)^2}{n}\right)V(t)
 -\frac{2\gmin(t)}{n-1}V(t)+q_{\max}\gmax(t)^2.
 \label{eq:joint-RS}
\end{equation}
\end{lemma}

The proof of \lemref{lem:joint-one-step} is in \appref{app:joint-convergence-proof}.

The following theorem makes the third contribution precise and records the moment assumptions needed beyond the Gaussian model.

\begin{theorem}[Stabilisation with distance-dependent joint noise]
\label{thm:effective-signed}
Fix a finite geometry, independent uniform ordered interactions, and local initiator gains \eqref{eq:gain} with $1/2<p\le1$. Suppose the effective errors satisfy
\begin{equation}
 \E[\eta_t\mid\F_t,I_t=i,R_t=r]=0,
 \qquad
 \E[\eta_t^2\mid\F_t,I_t=i,R_t=r]\le q_{\max}<\infty
 \label{eq:effective-assumptions}
\end{equation}
for every pair and time. Then the signed update \eqref{eq:joint-signed-x} admits a finite random variable $C_\infty$ such that
\begin{equation}
 x_i(t)\longrightarrow p_i+C_\infty
 \qquad\text{for every }i\quad\text{with probability 1}.
 \label{eq:effective-limit}
\end{equation}
In particular, the conclusion holds under the Gaussian joint model \eqref{eq:scenario-both}. A positive lower bound on the effective variance is not required.
\end{theorem}

The complete proof of \thmref{thm:effective-signed} is in \appref{app:joint-convergence-proof}. It applies Robbins--Siegmund to \eqref{eq:joint-RS} and uses the same translation-mean decomposition as in the fixed-noise signed protocol.

\begin{theorem}[Relative-coordinate rate under joint noise]
\label{thm:joint-relative}
Under the Gaussian joint model \eqref{eq:scenario-both}, or more generally under \eqref{eq:effective-assumptions} with a uniform conditional fourth-moment bound,
\begin{equation}
 \E V(t)=O(t^{-p}),\qquad
 \bigl\|(x_i(t)-x_j(t))-(p_i-p_j)\bigr\|_2=O(t^{-p/2}).
 \label{eq:joint-relative-rate}
\end{equation}
\end{theorem}

\begin{proof}
The proof is given in \lemref{lem:joint-ms-rate} in \appref{app:joint-ms-rate}. The argument combines \eqref{eq:joint-RS}, the counter-concentration event, the polynomial moment envelope, and \lemref{lem:rate-comparison}.
\end{proof}

The worst-pair variance $q_{\max}$ suffices for convergence and disagreement
bounds. The leading common-translation fluctuation admits a sharper
constant involving the \emph{average combined variance} $\bar q$. Write
\[
 \mathcal Z(t)=\sum_{s=0}^{t-1}\frac{\gamma_{I_s}(s)}n\eta_s
\]
for the cumulative noise contribution, and use the completion time
$\tau_k^{\rm loc}=\inf\{t:\min_i c_i(t)\ge k\}$ from
\thmref{thm:signed-rate}.

\begin{theorem}[Common-translation rate under joint noise]
\label{thm:joint-rate}
Under the Gaussian joint model of \thmref{thm:effective-signed},
\begin{equation}
 \bigl\|\mathcal Z(\infty)-\mathcal Z(\tau_k^{\rm loc})\bigr\|_2^2
 =\frac{\bar q}{n(2p-1)}k^{1-2p}(1+o(1)).
 \label{eq:joint-noise-tail}
\end{equation}
For every $0<\delta<\min\{p-1/2,p/2\}$,
\begin{equation}
 \bigl\|C_\infty-M(\tau_k^{\rm loc})\bigr\|_2
 \le\sqrt{\frac{\bar q}{n(2p-1)}}\,k^{1/2-p}(1+o(1))
 +O\!\left(k^{1/2-3p/2+\delta}\right).
 \label{eq:joint-common-tail}
\end{equation}
When $\bar q>0$, the full common-translation RMS tail is asymptotic to the
first term in \eqref{eq:joint-common-tail}. Conditional on the entire pair
schedule, the noise-tail variance is at most
\begin{equation}
 \frac{q_{\max}}n\sum_{m=k}^{\infty}(m+1)^{-2p}.
 \label{eq:joint-schedule-bound}
\end{equation}
\end{theorem}

The proof is in \appref{app:joint-rate-proof}. Thus allowing \emph{both}
error sources to vary with distance changes the geometry-dependent
constants, but not the cooling exponents for a fixed finite population.
The common translation remains the slower mode for $1/2<p<1$.

\subsection{Failure of \RightAlign{} under combined noise}
\label{subsec:joint-right-align}

Using the same received label and displacement measurement in \RightAlign{}
would instead give
\begin{equation}
 y_i^+=\max\{y_i,y_r+\eta_t\}.
 \label{eq:effective-max}
\end{equation}
Distance-dependent errors do not remove its rectification mechanism.

\begin{observation}[\RightAlign{} under distance-dependent joint noise]
\label{obs:joint-max}
Assume \eqref{eq:scenario-both} and $q_{\min}>0$. For the rule
\eqref{eq:effective-max},
\begin{equation}
 \E[H(t+1)-H(t)\mid\F_t]
 \ge\frac{\sqrt{q_{\min}}}{n\sqrt{2\pi}}.
 \label{eq:effective-max-drift}
\end{equation}
Consequently,
$\E H(t)\ge H(0)+t\sqrt{q_{\min}}/(n\sqrt{2\pi})$. Moreover, with probability 1,
\[
 H(t)\to+\infty,
 \qquad \limsup_{t\to\infty}R(t)=\infty.
\]
If $q_{ir}=d_{ir}$, the same failure holds, with drift at least
$\sqrt{d_{\min}}/(n\sqrt{2\pi})$.
\end{observation}

At consistency the mean one-sided increment for pair $(i,r)$ is
$\sqrt{q_{ir}/(2\pi)}>0$. A maximising responder occurs with probability at
least $1/n$, giving the drift bound; the full proof is in
\appref{app:joint-max-proof}. In contrast, \SignedUpdate{} uses a signed correction whose random component remains centred, even when both component variances grow with distance.

\subsection{Examples and scope}

The results above do not prescribe a unique setting. For example, both
communication and sensing variances may increase affinely with distance,
while conditional correlation changes the effective variance to
$q_{ir}=\Var(\zeta_t-\varepsilon_t\mid I_t=i,R_t=r)$. The signed
convergence theorem requires only conditional centring and a bounded
conditional second moment for the effective error. A complete affine
example, the correlated-error extension, model limitations, and
single-source specialisations are collected in
Appendices~\ref{app:joint-examples-scope}--\ref{app:single-source}.

\section{Conclusion and open directions}
\label{sec:conclusion}

The three contributions show how local-counter cooling stabilises increasingly expressive noisy coordination models. First, for averaging with a fixed centred finite-variance communication-error distribution, gains with $1/2<p\le1$ retain enough cumulative mixing to eliminate disagreement while making the accumulated noise convergent. Under these same assumptions, the relative RMS disagreement is $O(k^{-p/2})$, whereas the 
RMS distance of the population mean from its limiting value has order $k^{1/2-p}$ for positive noise variance. Unequal local gains contribute only lower-order corrections.

Second, for one-dimensional localisation with fixed non-degenerate Gaussian displacement errors, the form of the update matters as much as the gain. \RightAlign{} rectifies the noise, and power-law damping alone does not recover both noise robustness and noiseless correctness. \SignedUpdate{} preserves the centring of the effective error and converges to the true geometry up to one finite random translation, with the same two-mode rate structure.

Third, the joint model makes \emph{both} communication and displacement-measurement errors depend on physical interaction distance. For independent sources, the combined variance is $q_{ir}=s_c^2(d_{ir})+s_m^2(d_{ir})$. The signed protocol still converges for every fixed finite geometry. The worst-pair combined variance controls the general bounds, while its average determines the leading common-translation fluctuation. Thus the distance-dependent contributions of communication and sensing are both visible in the guarantees, without imposing a single hardware-specific variance law.

Open directions include calibrated communication and sensing models, distance-dependent interaction probabilities, sharp high-probability finite-time bounds, guarantees uniform in the population size, persistent bias and temporal dependence beyond conditional centring, packet loss and quantised channels, adaptive gains, and tracking under moving geometries. These require assumptions beyond the static, uniformly scheduled additive-noise model studied here.

\label{page:main-last}
\clearpage
\printbibliography[title={References}]

\section*{AI Disclosure}
OpenAI's ChatGPT was used to assist with language polishing, manuscript
organisation, proof revision and restructuring, \LaTeX{} editing, and
consistency checking. The authors independently verified all mathematical
statements, proofs, citations, and conclusions and take full responsibility
for the final manuscript.

\clearpage
\appendix

\section{Auxiliary probabilistic tools}
\label{app:prob-tools}

\subsection{Local-counter and gain control}
\label{app:counter-gain-proofs}

\begin{proof}[Proof of \lemref{lem:counter-concentration}]
Fix an agent $i$. In the averaging scheduler let $B_i(s)$ be the indicator that $i$ belongs to
the unordered pair selected at step $s$. Then $B_i(s)$ are i.i.d. Bernoulli$(2/n)$ and
$c_i(t)=\sum_{s=0}^{t-1}B_i(s)$. In the localisation scheduler, with $B_i(s)$ denoting the
event that $i$ is the initiator, the variables are i.i.d. Bernoulli$(1/n)$. Hence in either case
$c_i(t)$ is binomial with mean $\lambda t$.

Hoeffding's inequality for a sum of $t$ independent random variables in $[0,1]$ gives
\[
 \Pp(|c_i(t)-\lambda t|>t^{1/2+\delta})
 \le2\exp\!\left(-\frac{2t^{1+2\delta}}{t}\right)
 =2e^{-2t^{2\delta}}.
\]
Taking a union bound over the $n$ agents proves \eqref{eq:counter-tail}. Since
$\sum_{t\ge1}e^{-2t^{2\delta}}<\infty$, Borel--Cantelli implies that the event in
\eqref{eq:counter-tail} occurs only finitely often with probability 1. Because $n$ is fixed, the
conclusion holds simultaneously for every agent. This proves \eqref{eq:counterconc}.
\end{proof}

\begin{proof}[Proof of \lemref{lem:gain-control}]
Fix $0<\delta<\min\{1/2,p-1/2\}$ and work on the probability-one event from
\lemref{lem:counter-concentration}. Uniformly in $i$,
\[
 c_i(t)+1=\lambda t\left(1+O(t^{-1/2+\delta})\right).
\]
Since $u\mapsto u^{-p}$ is continuously differentiable on every interval bounded away from
zero,
\[
 (c_i(t)+1)^{-p}
 = (\lambda t)^{-p}\left(1+O(t^{-1/2+\delta})\right),
\]
which proves \eqref{eq:gain-asymptotic}.

For the mismatch, let $f(u)=(u+1)^{-p}$. On the same event, every counter lies between two
positive multiples of $t$ for all sufficiently large $t$, and therefore
\[
 \sup_{u\text{ between }c_i(t),c_j(t)}|f'(u)|
 =O(t^{-p-1}).
\]
By the mean-value theorem and \eqref{eq:counterconc},
\[
 |\gamma_i(t)-\gamma_j(t)|
 \le C t^{-p-1}|c_i(t)-c_j(t)|
 =O(t^{-p-1/2+\delta}),
\]
uniformly in $i,j$. Since $\delta<p-1/2$, the exponent is strictly below $-1$, so
$\sum_t\Dg(t)<\infty$.

Finally \eqref{eq:gain-asymptotic} yields $\gmin(t),\gmax(t)=\Theta(t^{-p})$ with probability 1.
Thus $\sum\gmin=\infty$ because $p\le1$, and $\sum\gmax^2<\infty$ because $2p>1$.
\end{proof}

\subsection{Martingale, weighted-event, moment, and rate lemmas}

\begin{proof}[Proof of \lemref{lem:mart-series}]
Let $S_m=\sum_{t=0}^{m-1}D_t$. For $r>s$, orthogonality of martingale differences gives
\[
 \E[(S_r-S_s)^2]=\sum_{t=s}^{r-1}\E[D_t^2].
\]
Hence if the unconditional sum of variances is finite, $(S_m)$ is Cauchy in $L^2$ and
converges in $L^2$ and with probability 1 by the martingale convergence theorem. The
tail identity follows by letting $r\to\infty$.

For the conditional-variance statement, put $v_t=\E[D_t^2\mid\F_t]$ and $A_t=\sum_{s=0}^t v_s$. For an integer $m\ge1$, define the predictably truncated differences
\[
 D_t^{(m)}=D_t\1_{\{A_t\le m\}}.
\]
The indicator is $\F_t$-measurable, so these are martingale differences. Moreover,
\[
 \sum_t\E[(D_t^{(m)})^2]
 =\E\!\left[\sum_t v_t\1_{\{A_t\le m\}}\right]\le m.
\]
Their series therefore converges with probability 1 by the unconditional statement already proved. On the event $\sum_t v_t<\infty$, there is an integer $m$ larger than this sum, and then $D_t^{(m)}=D_t$ for every $t$. Taking the countable intersection of the convergence events over $m$ proves the  
assertion holding with probability 1. This truncation excludes an increment before it could overshoot the variance budget.
\end{proof}

\begin{lemma}[Nonsummable weights see positive-density events]
\label{lem:weighted-bernoulli}
Let $B_k$ be independent Bernoulli variables with success probability $q>0$.
Let $a_k\ge0$ be deterministic and assume
\[
 \sum_k a_k=\infty,\qquad \sum_k a_k^2<\infty.
\]
Then
\begin{equation}
 \sum_{k=0}^\infty a_kB_k=\infty\qquad\text{with probability 1}
 \label{eq:weighted-bernoulli}
\end{equation}
\end{lemma}

\begin{proof}[Proof of \lemref{lem:weighted-bernoulli}]
Write
\[
 \sum_{k<K}a_kB_k
 =q\sum_{k<K}a_k+\sum_{k<K}a_k(B_k-q).
\]
The second term is a martingale with quadratic variation
$q(1-q)\sum_ka_k^2<\infty$. By \lemref{lem:mart-series} it converges with probability 1 to a finite
random limit. The first term tends to $+\infty$ because $q>0$ and $\sum_ka_k=\infty$.
Therefore the left-hand side diverges to $+\infty$ with probability 1.
\end{proof}

\begin{lemma}[Polynomial moment envelope]
\label{lem:moment-envelope}
For the signed localisation process with effective errors of uniformly bounded fourth
moment, there is a constant $C$ such that $\E[V(t)^2]\le C(1+t)^4$.
\end{lemma}

\begin{proof}
Put $S(t)=\max_i|y_i(t)|$ and write the effective error as $\eta_t$ (equal to $-\xi_t$ in
the fixed sensing-only model). Since $0\le a\le1$, the update
$y_i^+=(1-a)y_i+a y_r+a\eta_t$ gives
\[
 S(t+1)\le S(t)+|\eta_t|.
\]
By Minkowski's inequality in $L^4$ and the uniform fourth-moment bound,
$\|S(t)\|_4=O(1+t)$. Since $V(t)=\sum_i y_i(t)^2-nM(t)^2\le nS(t)^2$,
$\E[V(t)^2]\le n^2\E[S(t)^4]=O((1+t)^4)$.
\end{proof}

\begin{lemma}[Schedule-conditional second-moment envelope]
\label{lem:avg-conditional-envelope}
For averaging under the assumptions of \thmref{thm:avg}, let $\mathcal S$ be the sigma-field
generated by the entire pair schedule and put $S_0=\max_i|x_i(0)|$. Then
\begin{equation}
 \E[V(t)\mid\mathcal S]\le n(S_0+t\sigma)^2.
 \label{eq:avg-schedule-envelope}
\end{equation}
Consequently, for every $A\in\mathcal S$,
$\E[V(t)\1_A]\le n(S_0+t\sigma)^2\Pp(A)$.
\end{lemma}

\begin{proof}
Put $S(t)=\max_i|x_i(t)|$. The convex-combination part of each update gives
\[
 S(t+1)\le S(t)+\frac12\bigl(|\xi_i(t)|+|\xi_j(t)|\bigr).
\]
Conditionally on $\mathcal S$, the channel errors retain their independent fixed law with
second moment $\sigma^2$. Iterating this inequality and applying Minkowski's inequality in
the conditional $L^2$ norm yields
\[
 \bigl(\E[S(t)^2\mid\mathcal S]\bigr)^{1/2}\le S_0+t\sigma.
\]
Since $V(t)=\sum_i x_i(t)^2-nM(t)^2\le nS(t)^2$, this proves
\eqref{eq:avg-schedule-envelope}. Multiplying by $\1_A$ and taking expectations proves
the final assertion.
\end{proof}

\begin{proof}[Proof of \lemref{lem:rate-comparison}]
First take $1/2<p<1$. Choose $T$ so large that the $o(t^{-p})$ term in
\eqref{eq:rate-rec-general} has absolute value at most $(a/2)t^{-p}$ for $t\ge T$, and absorb
$r_t$ into $b't^{-2p}$ because it is superpolynomially small. Thus
\[
 u_{t+1}\le(1-a't^{-p})u_t+b't^{-2p},\qquad a'=a/2>0.
\]
Choose $K>b'/a'$. We claim $u_t\le Kt^{-p}$ for all sufficiently large $t$. If it holds at
$t$, then
\[
 u_{t+1}\le Kt^{-p}-(a'K-b')t^{-2p}.
\]
On the other hand
\[
 K(t+1)^{-p}=Kt^{-p}-pKt^{-p-1}+O(t^{-p-2}).
\]
Since $2p<p+1$ for $p<1$, the negative term of order $t^{-2p}$ eventually dominates the
required decrement of order $t^{-p-1}$. After increasing $K$ to cover finitely many initial
times, induction proves $u_t=O(t^{-p})$.

For $p=1$, write the coefficient as $1-(a+o(1))/t$ with $a>1$. Pick $a'$ with
$1<a'<a$ and $T$ so that the recursion is
\[
 u_{t+1}\le(1-a'/t)u_t+b'/t^2.
\]
Choose $K>b'/(a'-1)$. If $u_t\le K/t$, then
\[
 u_{t+1}\le \frac Kt-\frac{a'K-b'}{t^2}
 \le \frac Kt-\frac K{t^2}+O(t^{-3})
 \le \frac K{t+1}
\]
for all sufficiently large $t$. Again enlarge $K$ to start the induction. Hence
$u_t=O(t^{-1})$.
\end{proof}

\begin{lemma}[Uniform conditional occurrence]
\label{lem:conditional-occurrence}
Let $E_t\in\F_{t+1}$ and suppose $\Pp(E_t\mid\F_t)\ge q>0$ with probability 1 for every $t$.
Then $E_t$ occurs infinitely often with probability 1.
\end{lemma}

\begin{proof}
For any deterministic $s$ and $m\ge1$, repeated conditioning gives
\[
 \Pp(E_s^c\cap\cdots\cap E_{s+m-1}^c\mid\F_s)\le(1-q)^m.
\]
Letting $m\to\infty$ shows that, conditional on $\F_s$, the probability of no future occurrence
is zero. Taking a countable intersection over $s$ proves infinitely many occurrences with probability 1.
\end{proof}

\section{Averaging convergence with probability 1}
\label{app:avg-convergence-proof}

\begin{proof}[Proof of \lemref{lem:avg-one-step}]
Fix $t$ and condition on $\F_t$. Suppose $\{i,j\}$ is selected, set
$a=\gamma_i(t)$, $b=\gamma_j(t)$, and $d=z_i-z_j=x_i-x_j$. The two nonzero state increments
are
\[
 \Delta_i=\frac a2(-d+\xi_i),\qquad
 \Delta_j=\frac b2(d+\xi_j).
\]
Let $Q=I-n^{-1}\1\1^\top$, so $z=Qx$ and $V=\|z\|^2$. Since $Q$ is an orthogonal
projection,
\[
 V^+=\|z+Q\Delta\|^2
 \le V+2z^\top\Delta+\|\Delta\|^2.
\]
Using $\E\xi_i=\E\xi_j=0$, independence, and
$\E\xi_i^2=\E\xi_j^2=\sigma^2$,
\begin{align*}
 \E_\xi[2z^\top\Delta]
 &=-\frac{a+b}{2}d^2+\frac{b-a}{2}(z_i+z_j)d,\\
 \E_\xi\|\Delta\|^2
 &=\frac{a^2+b^2}{4}(d^2+\sigma^2).
\end{align*}
The asymmetry term obeys
\[
 \left|\frac{b-a}{2}(z_i+z_j)d\right|
 =\frac{|b-a|}{2}|z_i^2-z_j^2|
 \le\frac{\Dg(t)}2V.
\]
Also $a+b\ge2\gmin$ and $a^2+b^2\le2\gmax^2$. For every zero-sum vector $z$,
\[
 \sum_{i<j}(z_i-z_j)^2=n\sum_i z_i^2=nV,
\]
so a uniform unordered pair satisfies
\[
 \E_{\rm pair}[d^2\mid\F_t]=\frac{2}{n-1}V.
\]
Averaging the preceding inequality over the pair yields
\begin{equation}
 \E[V^+\mid\F_t]
 \le
 \left(1+\frac{\Dg}2+\frac{\gmax^2}{n-1}\right)V
 -\frac{2\gmin}{n-1}V
 +\frac{\sigma^2}{2}\gmax^2.
 \label{eq:avg-RS}
\end{equation}
\end{proof}

\begin{proof}[Proof of the disagreement part of \thmref{thm:avg}]
Apply \lemref{lem:RS} to \eqref{eq:avg-RS} with
\[
 A_t=\frac{\Dg(t)}2+\frac{\gmax(t)^2}{n-1},\quad
 B_t=\frac{2\gmin(t)}{n-1}V(t),\quad
 C_t=\frac{\sigma^2}{2}\gmax(t)^2.
\]
By \lemref{lem:gain-control}, $\sum A_t<\infty$ and $\sum C_t<\infty$ with probability 1. Hence
$V(t)$ converges with probability 1 to a finite random limit $V_\infty$ and
\[
 \sum_t\gmin(t)V(t)<\infty\qquad\text{with probability 1}
\]
If $V_\infty>0$, then eventually $V(t)\ge V_\infty/2$, so
$\sum_t\gmin(t)V(t)=\infty$ because $\sum_t\gmin(t)=\infty$, a contradiction. Therefore
$V(t)\to0$ with probability 1.
\end{proof}

\begin{proof}[Completion of the proof of \thmref{thm:avg}]
For the selected pair, summing the two increments and dividing by $n$ gives
\[
 M(t+1)-M(t)=D_t+Z_t,
\]
where
\[
 D_t=\frac{(b-a)(x_i-x_j)}{2n},\qquad
 Z_t=\frac{a\xi_i+b\xi_j}{2n}.
\]
Since $|x_i-x_j|=|z_i-z_j|\le\sqrt{2V(t)}$,
\[
 |D_t|\le\frac{\Dg(t)}{\sqrt2\,n}\sqrt{V(t)}.
\]
The convergence of $V(t)$ established above implies $\sup_tV(t)<\infty$ with probability 1, and
\lemref{lem:gain-control} gives $\sum_t\Dg(t)<\infty$. Thus
$\sum_t|D_t|<\infty$ with probability 1.

The noise increment $Z_t$ is a martingale difference. Write
\[
 \mathcal Z(t)=\sum_{s=0}^{t-1}Z_s
\]
for its cumulative contribution to the mean. Conditionally on $\F_t$ and the selected pair,
\[
 \E[Z_t^2\mid\F_t,i,j]
 =\frac{\sigma^2}{4n^2}(a^2+b^2)
 \le\frac{\sigma^2}{2n^2}\gmax(t)^2.
\]
The conditional variances are summable with probability 1 by \lemref{lem:gain-control}, so
\lemref{lem:mart-series} implies that $\sum_tZ_t$ converges with probability 1. Therefore $M(t)$
converges to a finite random variable $X_\infty$. Finally
$|x_i(t)-M(t)|^2\le V(t)\to0$, so $x_i(t)\to X_\infty$ for every $i$ simultaneously. This
proves \thmref{thm:avg}.
\end{proof}

\section{Averaging rate proof}
\label{app:avg-rate-proof}

\subsection{Mean-square disagreement}

\begin{lemma}[Averaging disagreement rate]
\label{lem:avg-ms-rate}
Under the assumptions of \thmref{thm:avg-rate},
\[
 \E V(t)=O(t^{-p}),
 \qquad
 \bigl\|x_i(t)-x_j(t)\bigr\|_2=O(t^{-p/2})
\]
for every pair $i,j$.
\end{lemma}

\begin{proof}
Fix $0<\delta<p-1/2$ and define the good counter event
\[
 G_t=\left\{\max_i|c_i(t)-(2/n)t|\le t^{1/2+\delta}\right\}.
\]
By \eqref{eq:counter-tail}, $\Pp(G_t^c)\le2n e^{-2t^{2\delta}}$. On $G_t$,
\begin{align*}
 \gmin(t)&=(n/2)^pt^{-p}(1+O(t^{-1/2+\delta})),\\
 \gmax(t)^2&=(n/2)^{2p}t^{-2p}(1+O(t^{-1/2+\delta})),\\
 \Dg(t)&=O(t^{-p-1/2+\delta}).
\end{align*}
Substitution into \eqref{eq:avg-RS} yields, on $G_t$,
\begin{equation}
 \E[V(t+1)\mid\F_t]
 \le\left(1-a_{n,p}t^{-p}+o(t^{-p})\right)V(t)+b_{n,p}t^{-2p},
 \label{eq:avg-good-rec}
\end{equation}
where
\[
 a_{n,p}=\frac{2(n/2)^p}{n-1}>0.
\]
For $p=1$, $a_{n,1}=n/(n-1)>1$.

Outside $G_t$ we only need a crude polynomial bound. From the proof of
\lemref{lem:avg-one-step}, using $0\le a,b\le1$ and $(u-v)^2\le2u^2+2v^2$, there are constants
$C_0,C_1$ depending only on $n$ and the noise variance such that
\[
 \E[V(t+1)\mid\F_t]\le C_0V(t)+C_1.
\]
The bad-counter event $G_t^c$ depends only on the pair schedule, so
\lemref{lem:avg-conditional-envelope} gives
\[
 \E[V(t)\1_{G_t^c}]
 \le n(S_0+t\sigma)^2\Pp(G_t^c)
 \le C(1+t)^2e^{-2t^{2\delta}},
\]
which is smaller than every inverse polynomial. Let $u_t=\E V(t)$. Splitting expectation over
$G_t$ and $G_t^c$ in \eqref{eq:avg-good-rec} therefore gives
\[
 u_{t+1}
 \le(1-a_{n,p}t^{-p}+o(t^{-p}))u_t+O(t^{-2p})+r_t,
\]
with superpolynomially decreasing $r_t$. The conclusion $u_t=O(t^{-p})$ now follows from
\lemref{lem:rate-comparison}; the endpoint $p=1$ is valid because $a_{n,1}>1$.
Since $(x_i-x_j)^2=(z_i-z_j)^2\le2V$, the relative RMS bound follows.
\end{proof}

\subsection{RMS distance of the population mean from its limiting value}

\begin{lemma}[]
\label{lem:avg-common-tail}
Under the assumptions of \thmref{thm:avg-rate}, for every fixed
$0<\delta<\min\{p-1/2,p/2\}$,
\[
 \|M(\infty)-M(\tau_k^{\rm av})\|_2
 \le
 \frac{\sigma}{2\sqrt{n(2p-1)}}k^{1/2-p}(1+o(1))
 +O\!\left(k^{1/2-3p/2+\delta}\right).
\]
If $\sigma>0$, the full RMS distance of the population mean from its limiting value has order $k^{1/2-p}$.
\end{lemma}

\begin{proof}
Use the decomposition $M(t+1)-M(t)=D_t+Z_t$ from the proof of
\thmref{thm:avg}. Consider first the noise martingale $\mathcal Z$. At time
$\tau_k^{\rm av}$ every agent has already participated at least $k$ times. Conditional on the
entire pair schedule, the future error attached to a participation of any fixed agent
with pre-interaction counter $m$ appears in the population mean with coefficient
\[
 \frac{(m+1)^{-p}}{2n}.
\]
There are $n$ agents and the channel errors are independent, so the conditional variance of the future noise sum is at most
\[
 \frac{\sigma^2}{4n}\sum_{m=k}^{\infty}(m+1)^{-2p}.
\]
The same bound at $k=0$ is finite, so $\mathcal Z$ converges in $L^2$.
By \lemref{lem:mart-series} and \eqref{eq:ptail},
\begin{equation}
 \|\mathcal Z(\infty)-\mathcal Z(\tau_k^{\rm av})\|_2
 \le\frac{\sigma}{2\sqrt{n(2p-1)}}k^{1/2-p}(1+o(1)).
 \label{eq:avg-gaussian-tail-app}
\end{equation}

For completeness, the leading order is exact when $\sigma>0$. Counter concentration gives $\tau_k^{\rm av}\sim k/\lambda_{\rm av}$ and
\[
 \frac{c_i(\tau_k^{\rm av})}{k}\longrightarrow1
 \qquad\text{with probability 1, for every }i.
\]
Indeed, $\min_i c_i(\tau_k^{\rm av})=k$, and all counters are asymptotic to $\lambda_{\rm av}t$. Each agent's future contributions start at its actual counter, so orthogonality gives
\begin{align*}
 \|\mathcal Z(\infty)-\mathcal Z(\tau_k^{\rm av})\|_2^2
 &=\frac{\sigma^2}{4n^2}\E\!\left[
   \sum_{i=1}^n\sum_{m=c_i(\tau_k^{\rm av})}^{\infty}(m+1)^{-2p}\right]\\
 &=\frac{\sigma^2}{4n(2p-1)}k^{1-2p}(1+o(1)).
\end{align*}
The last step follows from \eqref{eq:ptail} and dominated convergence: every inner sum starts at a counter at least $k$, and its normalised value tends with probability 1 to the same limit. This argument uses independence and second moments, not Gaussianity. For Gaussian errors, the noise sum is Gaussian conditional on the pair schedule; no unconditional Gaussian law for the full consensus limit is asserted.

It remains to show that the gain-mismatch tail is smaller. Use the good-counter events
$G_t$ from the proof of \lemref{lem:avg-ms-rate}, with the present value of $\delta$. On $G_t$,
\[
 D_t^2\le C\Dg(t)^2V(t),
\]
and therefore, by \lemref{lem:avg-ms-rate},
\[
 \E[D_t^2\1_{G_t}]
 =O(t^{-2p-1+2\delta})O(t^{-p})
 =O(t^{-3p-1+2\delta}).
\]
On $G_t^c$, use $\Dg(t)\le1$ and \lemref{lem:avg-conditional-envelope} to obtain
\[
 \E[D_t^2\1_{G_t^c}]
 \le C\E[V(t)\1_{G_t^c}]
 \le C(1+t)^2e^{-2t^{2\delta}},
\]
which is superpolynomially small. Hence
\[
 \|D_t\|_2=O(t^{-3p/2-1/2+\delta}).
\]
These norms are summable because $\delta<p-1/2$. Thus $\sum_t D_t$ converges in $L^2$;
together with the noise-martingale convergence, this proves $M(t)\to M(\infty)$ in $L^2$
with the same limit holding with probability 1 as in \thmref{thm:avg}. Minkowski's inequality gives
\[
 \left\|\sum_{s\ge t}D_s\right\|_2
 \le\sum_{s\ge t}\|D_s\|_2
 =O(t^{1/2-3p/2+\delta}).
\]
Since $\tau_k^{\rm av}\ge k$ deterministically, Minkowski's inequality also gives
\begin{align*}
 \left\|\sum_{s\ge\tau_k^{\rm av}}D_s\right\|_2
 &\le \sum_{s\ge k}\|D_s\1_{\{\tau_k^{\rm av}\le s\}}\|_2\\
 &\le \sum_{s\ge k}\|D_s\|_2
 =O(k^{1/2-3p/2+\delta}).
\end{align*}
Combining this with \eqref{eq:avg-gaussian-tail-app} proves \eqref{eq:avg-commonrate}. Since $\delta<p/2$, the correction is $o(k^{1/2-p})$. The reverse triangle inequality and the noise-tail asymptotic therefore also give the stated order of the full 
RMS distance of the population mean from its limiting value
when $\sigma>0$.
\end{proof}

\begin{proof}[Proof of \thmref{thm:avg-rate}]
The mean-square and relative-disagreement claims are \lemref{lem:avg-ms-rate}; the common-mode claim is \lemref{lem:avg-common-tail}. Rewriting sequential time in the deterministic local-count scale $k=(2/n)t$ changes only constants and not exponents.
\end{proof}

\subsection{Rate-exponent summary}

\begin{table}[H]
\centering
\caption{Late-stage exponents for local power-law cooling. The full RMS distance column gives the RMS order for positive noise variance; the relative-disagreement column gives an RMS upper-bound rate. Constants may depend on the fixed population size.}
\label{tab:p}
\begin{tabular}{@{}ccc@{}}
\toprule
$p$ & full RMS distance & Relative-disagreement RMS bound \\
\midrule
$0.55$ & $k^{-0.05}$ & $k^{-0.275}$ \\
$0.75$ & $k^{-0.25}$ & $k^{-0.375}$ \\
$0.90$ & $k^{-0.40}$ & $k^{-0.45}$ \\
$1$ & $k^{-1/2}$ & $k^{-1/2}$ \\
\bottomrule
\end{tabular}
\end{table}

\section{\RightAlign{} non-convergence and the damping obstruction}
\label{app:right-align}

This appendix contains the detailed arguments underlying the observations in \secref{sec:localisation}. The first subsection proves the non-convergence observation; the second explains why power-law damping alone does not restore the required localisation guarantee.

\subsection{Plain \RightAlign{}: proof of non-convergence}
\label{app:right-align-plain}

\begin{proof}[Proof of \obsref{obs:maxfail}]
The process $H(t)=\max_i y_i(t)$ is nondecreasing. Conditional on $\F_t$, choose any agent
$r$ with $y_r(t)=H(t)$. Under the uniform ordered scheduler, the responder is uniform on the
$n$ agents, so the probability that the actual responder is a current maximiser is at least
$1/n$. On that event, if $\eta_t>0$, the candidate value offered to the initiator is
$H(t)+\eta_t$, hence
\[
 H(t+1)-H(t)\ge[\eta_t]_+.
\]
Therefore
\[
 \E[H(t+1)-H(t)\mid\F_t]
 \ge\frac1n\E\!\left[\pos{\eta_t}\right]
 =\frac{\sigma}{n\sqrt{2\pi}},
\]
proving \eqref{eq:maxdrift}. Taking expectations and summing over time gives the stated linear
lower bound on $\E H(t)$.

For the statements holding with probability 1 fix $A>0$ and let $E_t(A)$ be the event that the responder is
a current maximiser and $\eta_t>A$. Fresh Gaussian noise and the uniform scheduler imply
\[
 \Pp(E_t(A)\mid\F_t)
 \ge \frac1n\Pp(N(0,\sigma^2)>A)=:q_A>0.
\]
By \lemref{lem:conditional-occurrence}, $E_t(A)$ occurs infinitely often with probability 1. Each
occurrence raises $H$ by more than $A$, so in particular $H(t)\to+\infty$. Immediately after
such an occurrence, the responder is still at the old maximum while the updated initiator is
more than $A$ above it; hence $R(t+1)>A$. Applying the argument to every positive integer
$A$ and intersecting the resulting probability-one events proves
$\limsup_tR(t)=\infty$.
\end{proof}

The mechanism is rectification: at exact consistency the noise-induced increment is
$[\eta_t]_+$ and
\begin{equation}
 \E\!\left[\pos{\eta_t}\right]=\frac{\sigma}{\sqrt{2\pi}}>0.
 \label{eq:positivepart}
\end{equation}
\subsection{Why damping \RightAlign{} is insufficient}
\label{app:right-align-damping}
We now prove the two damping observations stated in the main narrative for the rule \eqref{eq:dampedmax}.

\begin{proof}[Proof of \obsref{obs:damped-max}]
Assume for contradiction that the event
\[
 \mathcal C=\{\exists C\in\R:\ y_i(t)\to C\text{ for all }i\}
\]
has positive probability. Fix an agent $i$. Under the uniform ordered scheduler, $i$ initiates
infinitely often with probability 1; index its initiations by $k=0,1,2,\ldots$. On $\mathcal C$ there
is a finite random $K$ such that at every initiation $k\ge K$ and for every possible responder
$r$,
\[
 y_r-y_i\ge-1.
\]
Let $B_k=\1_{\{\eta_{i,k}\ge2\}}$, where $\eta_{i,k}$ is the fresh error at the initiation with local index $k$. These indicators are i.i.d. Bernoulli with
$q=\Pp(N(0,\sigma^2)\ge2)>0$. Whenever $k\ge K$ and $B_k=1$,
\[
 [y_r-y_i+\eta_{i,k}]_+\ge1,
\]
so the increment of $y_i$ is at least $\gamma_k=(k+1)^{-p}$. Let $s_{i,k}$ be the sequential time of the initiation with local index $k$. The updates are nonnegative, so for every $L>K$, on $\mathcal C$,
\[
 y_i(s_{i,L}+1)-y_i(s_{i,K})
 \ge\sum_{k=K}^{L}\gamma_kB_k.
\]
For $1/2<p\le1$, $\sum\gamma_k=\infty$ and $\sum\gamma_k^2<\infty$, so
\lemref{lem:weighted-bernoulli} makes the right-hand side diverge as $L\to\infty$, with probability 1. This contradicts the assumed finite limit $C$. Hence $\Pp(\mathcal C)=0$.
\end{proof}

Conversely, choosing $p>1$ makes $\sum_k\gamma_k<\infty$. The following counterexample shows that this can destroy convergence to a common translation even without noise.

\begin{proof}[Proof of \obsref{obs:summable-max-counterexample}]
Take three agents with noiseless translation errors
$y_1(0)=1$ and $y_2(0)=y_3(0)=0$. Agent $1$ is a permanent maximiser because damped \RightAlign{} 
never decreases a value and, without noise, no value can exceed the current maximum $1$.
Condition on the event that the first interaction is $(2,3)$, which has probability $1/6$. Then $y_2=y_3=0$ at that interaction, so $y_2$ does not change and its local counter becomes $1$.

At every later initiation $k\ge1$ of agent $2$, regardless of the responder,
\[
 [y_r-y_2]_+\le1-y_2.
\]
Writing $g_k=1-y_2$ immediately before that initiation,
\[
 g_{k+1}\ge(1-\gamma_k)g_k.
\]
Since $p>1$, $\sum_{k\ge1}\gamma_k<\infty$, and $0<\gamma_k<1$ for every $k\ge1$. The
infinite product $\prod_{k\ge1}(1-\gamma_k)$ is therefore strictly positive. Consequently
\[
 \liminf_{k\to\infty}g_k
 \ge g_1\prod_{k\ge1}(1-\gamma_k)>0.
\]
Thus agent $2$ remains bounded away from the permanent maximum $1$, so convergence to a common translation fails. The conditioning event has positive probability, proving the observation.
\end{proof}

Together, Observations~\ref{obs:damped-max} and~\ref{obs:summable-max-counterexample} identify the obstruction within this power-law family. Nonsummable damping fails to suppress accumulated rectification under persistent Gaussian noise, whereas summable damping can leave a permanent noiseless discrepancy. \SignedUpdate{} avoids this first-order bias while retaining nonsummable total correction.

\section{Signed-localisation convergence with probability 1}
\label{app:signed-convergence-proof}

\begin{proof}[Proof of \lemref{lem:loc-one-step}]
Condition on $\F_t$ and suppose the ordered pair $(i,r)$ is selected. Write
$a=\gamma_i(t)$ and $d=z_i-z_r=y_i-y_r$. The only nonzero state increment is
\[
 \Delta y=a(-d-\xi_t)e_i.
\]
With $Q=I-n^{-1}\1\1^\top$,
\[
 z^+=z+a(-d-\xi_t)Qe_i.
\]
Since $\|Qe_i\|^2=1-1/n$ and $z^\top Qe_i=z_i$,
\[
 V^+=V+2a(-d-\xi_t)z_i
 +a^2(d+\xi_t)^2\left(1-\frac1n\right).
\]
Averaging over the fresh noise gives
\[
 \E_\xi[V^+]
 =V-2ad z_i+a^2(d^2+\sigma^2)\left(1-\frac1n\right).
\]
For a zero-sum vector $z$,
\[
 \sum_{r\ne i}(z_i-z_r)=nz_i,
 \qquad
 \sum_{r\ne i}(z_i-z_r)^2=nz_i^2+V.
\]
Therefore the ordered-pair average of the linear term is
\[
 -\frac{2}{n(n-1)}\sum_i\gamma_i z_i
 \sum_{r\ne i}(z_i-z_r)
 =-\frac{2}{n-1}\sum_i\gamma_i z_i^2
 \le-\frac{2\gmin}{n-1}V.
\]
For the quadratic $d^2$ term,
\begin{align*}
 &\frac{1-1/n}{n(n-1)}
 \sum_{i\ne r}\gamma_i^2(z_i-z_r)^2\\
 &\qquad=
 \frac{1-1/n}{n(n-1)}
 \left(n\sum_i\gamma_i^2z_i^2+\Bigl(\sum_i\gamma_i^2\Bigr)V\right)
 \le\frac{2}{n}\gmax^2V.
\end{align*}
The noise-variance contribution equals
\[
 \frac{1-1/n}{n(n-1)}\sum_{i\ne r}\gamma_i^2\sigma^2
 \le\sigma^2\gmax^2.
\]
Combining the three bounds gives
\begin{equation}
 \E[V(t+1)\mid\F_t]
 \le
 \left(1+\frac{2\gmax(t)^2}{n}\right)V(t)
 -\frac{2\gmin(t)}{n-1}V(t)
 +\sigma^2\gmax(t)^2.
 \label{eq:loc-RS}
\end{equation}
\end{proof}

\begin{proof}[Disagreement part of \thmref{thm:signed}]
Apply \lemref{lem:RS} to \eqref{eq:loc-RS} with
\[
 A_t=\frac2n\gmax(t)^2,
 \quad B_t=\frac{2\gmin(t)}{n-1}V(t),
 \quad C_t=\sigma^2\gmax(t)^2.
\]
By \lemref{lem:gain-control}, both $\sum A_t$ and $\sum C_t$ are finite with probability 1.
Hence $V(t)$ converges and
\[
 \sum_t\gmin(t)V(t)<\infty.
\]
Since $\sum_t\gmin(t)=\infty$, the limiting value of $V$ must be zero.
\end{proof}

\label{subsubsec:signed-mean-proof-main}
\begin{proof}[Completion of the proof of \thmref{thm:signed}]
For the selected ordered pair $(i,r)$, averaging the single state increment over the population
gives
\[
 M(t+1)-M(t)=G_t+Z_t,
\]
with
\[
 G_t=\frac{\gamma_i(t)}n(z_r-z_i),
 \qquad
 Z_t=-\frac{\gamma_i(t)}n\xi_t.
\]
As in the averaging proof, let $\mathcal Z(t)=\sum_{s<t}Z_s$ denote the cumulative noise contribution. Its increments are martingale differences and
\[
 \E[Z_t^2\mid\F_t,i,r]
 =\frac{\sigma^2\gamma_i(t)^2}{n^2}
 \le\frac{\sigma^2}{n^2}\gmax(t)^2.
\]
Thus $\sum_tZ_t$ converges with probability 1 by \lemref{lem:mart-series}.

Now decompose the pair-selection term as
$G_t=\mu_t+U_t$, where $\mu_t=\E[G_t\mid\F_t]$ and $U_t$ is the corresponding
martingale difference. Uniform averaging over ordered pairs gives
\begin{align*}
 \mu_t
 &=\frac1{n}\frac1{n(n-1)}
   \sum_{i\ne r}\gamma_i(z_r-z_i)\\
 &=-\frac1{n(n-1)}\sum_i\gamma_i z_i.
\end{align*}
Let $\bar\gamma=n^{-1}\sum_i\gamma_i$. Since $\sum_i z_i=0$,
\[
 |\mu_t|
 =\frac1{n(n-1)}\left|\sum_i(\gamma_i-\bar\gamma)z_i\right|
 \le C_n\Dg(t)\sqrt{V(t)}.
\]
The convergence of $V$ makes $\sup_tV(t)<\infty$ with probability 1, and
$\sum_t\Dg(t)<\infty$ by \lemref{lem:gain-control}; hence $\sum_t|\mu_t|<\infty$ with probability 1.

Finally,
\[
 \E[U_t^2\mid\F_t]
 \le\E[G_t^2\mid\F_t]
 \le\frac{\gmax(t)^2}{n^2}
 \E_{\rm pair}[(z_r-z_i)^2\mid\F_t]
 =\frac{2\gmax(t)^2}{n^2(n-1)}V(t).
\]
The right-hand side is summable with probability 1 because $V$ is bounded and
$\sum\gmax^2<\infty$. Hence $\sum_tU_t$ converges with probability 1 by
\lemref{lem:mart-series}. We have expressed $M(t)$ as its initial value plus one absolutely
summable predictable series and two convergent martingale series, so $M(t)\to C_\infty$ with probability 1. Since $V(t)\to0$, every $y_i(t)-M(t)\to0$ and therefore
$x_i(t)=p_i+y_i(t)\to p_i+C_\infty$. This completes the proof.
\end{proof}

\section{Signed-localisation rate proof}
\label{app:signed-rate-proof}

\subsection{Mean-square disagreement}

\begin{lemma}[Signed-localisation disagreement rate]
\label{lem:signed-ms-rate}
Under the assumptions of \thmref{thm:signed-rate},
\[
 \E V(t)=O(t^{-p}),
 \qquad
 \bigl\|(x_i(t)-x_j(t))-(p_i-p_j)\bigr\|_2=O(t^{-p/2}).
\]
\end{lemma}

\begin{proof}
Fix $0<\delta<p-1/2$ and define
\[
 G_t=\left\{\max_i|c_i(t)-t/n|\le t^{1/2+\delta}\right\}.
\]
On $G_t$,
\[
 \gmin(t)=n^pt^{-p}(1+O(t^{-1/2+\delta})),
 \qquad
 \gmax(t)^2=n^{2p}t^{-2p}(1+O(t^{-1/2+\delta})).
\]
Substituting into \eqref{eq:loc-RS} gives
\[
 \E[V(t+1)\mid\F_t]
 \le(1-a_{n,p}^{\rm loc}t^{-p}+o(t^{-p}))V(t)+O(t^{-2p})
\]
on $G_t$, where
\[
 a_{n,p}^{\rm loc}=\frac{2n^p}{n-1}>0,
 \qquad a_{n,1}^{\rm loc}=\frac{2n}{n-1}>1.
\]
The complement of $G_t$ has probability at most $2ne^{-2t^{2\delta}}$. By
\lemref{lem:moment-envelope} and Cauchy--Schwarz,
\[
 \E[V(t)\1_{G_t^c}]
 \le\|V(t)\|_2\Pp(G_t^c)^{1/2}
 \le C(1+t)^2e^{-t^{2\delta}}.
\]
Since \eqref{eq:loc-RS} also gives
$\E[V(t+1)\mid\F_t]\le C_0V(t)+C_1$, the bad-counter contribution is
superpolynomially small. Thus
$u_t=\E V(t)$ satisfies \eqref{eq:rate-rec-general} with the above coefficient, and
\lemref{lem:rate-comparison} yields $\E V(t)=O(t^{-p})$. Finally
\[
 ((x_i-x_j)-(p_i-p_j))^2=(z_i-z_j)^2\le2V,
\]
proving the relative RMS bound.
\end{proof}

\subsection{Common-translation tail}

\begin{lemma}[Signed-localisation common-translation tail]
\label{lem:signed-common-tail}
Under the assumptions of \thmref{thm:signed-rate}, for every fixed
$0<\delta<\min\{p-1/2,p/2\}$,
\[
 \|C_\infty-M(\tau_k^{\rm loc})\|_2
 \le
 \frac{\sigma}{\sqrt{n(2p-1)}}k^{1/2-p}(1+o(1))
 +O\!\left(k^{1/2-3p/2+\delta}\right).
\]
The noise-martingale tail is asymptotic to the first term, and the full common-translation RMS tail has order $k^{1/2-p}$.
\end{lemma}

\begin{proof}
Use the decomposition from \appref{app:signed-convergence-proof}:
\[
 C_\infty-M(t)=\sum_{s\ge t}\mu_s+\sum_{s\ge t}U_s+\sum_{s\ge t}Z_s.
\]
At $\tau_k^{\rm loc}$ every agent has initiated at least $k$ times. Conditional on the entire
pair schedule, the noise at the initiation of agent $i$ with local index $m$ enters the mean with
coefficient $(m+1)^{-p}/n$. Hence
\[
 \|\mathcal Z(\infty)-\mathcal Z(\tau_k^{\rm loc})\|_2^2
 \le\frac{\sigma^2}{n}\sum_{m=k}^{\infty}(m+1)^{-2p},
\]
which, by \eqref{eq:ptail}, gives the first term in \eqref{eq:loccommonrate}. As in \appref{app:avg-rate-proof}, the actual counters satisfy $c_i(\tau_k^{\rm loc})/k\to1$ with probability 1. Summing the variance from these actual counters and applying dominated convergence therefore gives
\[
 \|\mathcal Z(\infty)-\mathcal Z(\tau_k^{\rm loc})\|_2^2
 =\frac{\sigma^2}{n(2p-1)}k^{1-2p}(1+o(1)).
\]

For the pair-selection martingale, the estimate in \appref{app:signed-convergence-proof} and the
mean-square disagreement rate give
\[
 \E[U_t^2\1_{G_t}]=O(t^{-2p})O(t^{-p})=O(t^{-3p}).
\]
The complement contributes a superpolynomially small term by the moment envelope, so the same order holds for $\E[U_t^2]$. Orthogonality of martingale differences
therefore gives
\[
 \left\|\sum_{s\ge t}U_s\right\|_2
 =O(t^{1/2-3p/2}).
\]
For the predictable term, $|\mu_t|\le C\Dg(t)\sqrt{V(t)}$. Splitting over $G_t$ and its complement gives
\[
 \|\mu_t\|_2
 =O(t^{-p-1/2+\delta})O(t^{-p/2})
 =O(t^{-3p/2-1/2+\delta}),
\]
where the bad-event contribution is again superpolynomially small. Minkowski's inequality yields
\[
 \left\|\sum_{s\ge t}\mu_s\right\|_2
 =O(t^{1/2-3p/2+\delta}).
\]
Because $\tau_k^{\rm loc}\ge k$ deterministically, the same tail estimates remain valid at the
random time $\tau_k^{\rm loc}$: for the predictable series use Minkowski's inequality, and for the
pair-selection martingale use that $\1_{\{\tau_k^{\rm loc}\le s\}}$ is $\F_s$-measurable, so the tail increments remain orthogonal and their variances are bounded by the sum from $s=k$ onward. Thus both auxiliary
tails are $O(k^{1/2-3p/2+\delta})$, strictly smaller than the Gaussian $k^{1/2-p}$ term.
This proves the displayed bound. The reverse triangle inequality, the exact noise-tail asymptotic, and $\delta<p/2$ give the stated RMS order of the full common-translation tail.
\end{proof}

\begin{proof}[Proof of \thmref{thm:signed-rate}]
The relative-coordinate estimates follow from \lemref{lem:signed-ms-rate}. The common-translation estimate follows from \lemref{lem:signed-common-tail}. Passing from sequential time to the deterministic local-count scale $k=t/n$ changes only constants.
\end{proof}

\section{Convergence under joint distance-dependent noise with probability 1}
\label{app:joint-convergence-proof}

\begin{proof}[Proof of \lemref{lem:joint-one-step}]
Condition on $\F_t$ and on the ordered pair $(i,r)$. Put $a=\gamma_i(t)$ and $d=z_i-z_r=y_i-y_r$. The state increment is
\[
 \Delta y=a(-d+\eta_t)e_i.
\]
With $Q=I-n^{-1}\1\1^\top$,
\[
 V(t+1)=V(t)+2a(-d+\eta_t)z_i
       +a^2(-d+\eta_t)^2\left(1-\frac1n\right).
\]
Conditional centring removes the linear noise term, while the second-moment bound gives
\[
 \E[V(t+1)\mid\F_t,i,r]
 \le V(t)-2ad z_i+a^2(d^2+q_{\max})\left(1-\frac1n\right).
\]
The ordered-pair identities
\[
 \sum_{r\ne i}(z_i-z_r)=nz_i,
 \qquad
 \sum_{r\ne i}(z_i-z_r)^2=nz_i^2+V
\]
are the same as in \lemref{lem:loc-one-step}. Averaging over $(i,r)$ therefore bounds the linear term by $-2\gmin V/(n-1)$, the quadratic disagreement term by $2\gmax^2V/n$, and the noise term by $q_{\max}\gmax^2$. Combining the three bounds proves \eqref{eq:joint-RS}.
\end{proof}

\begin{proof}[Proof of \thmref{thm:effective-signed}]
Apply \lemref{lem:RS} to \eqref{eq:joint-RS}. By \lemref{lem:gain-control}, the positive error terms are summable and the total coefficient of $V(t)$ in the dissipative term is nonsummable. Hence $V(t)$ converges with probability 1 and
\[
 \sum_t\gmin(t)V(t)<\infty.
\]
Because $\sum_t\gmin(t)=\infty$, the limit must be zero.

For the mean translation, write
\[
 M(t+1)-M(t)=G_t+Z_t,
 \qquad
 G_t=\frac{\gamma_{I_t}(t)}n(z_{R_t}-z_{I_t}),
 \qquad
 Z_t=\frac{\gamma_{I_t}(t)}n\eta_t.
\]
The sequence $(Z_t)$ is a martingale difference and
\[
 \E[Z_t^2\mid\F_t]\le \frac{q_{\max}}{n^2}\gmax(t)^2,
\]
so $\sum_tZ_t$ converges with probability 1 by \lemref{lem:mart-series}. Decompose $G_t=\mu_t+U_t$, where $\mu_t=\E[G_t\mid\F_t]$. Uniform ordered-pair averaging and $\sum_i z_i=0$ give
\[
 \mu_t=-\frac1{n(n-1)}\sum_i\gamma_i(t)z_i(t),
 \qquad
 |\mu_t|\le C_n\Dg(t)\sqrt{V(t)}.
\]
Thus $\sum_t|\mu_t|<\infty$ with probability 1. Moreover,
\[
 \E[U_t^2\mid\F_t]
 \le \frac{2\gmax(t)^2}{n^2(n-1)}V(t),
\]
which is summable with probability 1 because $V(t)$ is bounded and $\sum_t\gmax(t)^2<\infty$. Hence $\sum_tU_t$ converges with probability 1. Therefore $M(t)$ has a finite limit $C_\infty$; together with $V(t)\to0$, this proves \eqref{eq:effective-limit}.
\end{proof}

\section{Distance-dependent combined errors: proofs and special cases}
\label{app:scenario3-proofs}
\label{app:joint-proofs}

This appendix supports the joint model of \secref{sec:physical-scenarios}.
The variance $q_{ir}$ includes both communication and displacement-measurement
errors. Single-source models are recovered only after the joint results.

\subsection{\RightAlign{} with combined distance-dependent errors}
\label{app:joint-max-proof}

\begin{proof}[Proof of \obsref{obs:joint-max}]
Condition on $\F_t$. The maximum $H(t)$ is nondecreasing. When the
responder is a current maximiser, the candidate offered to the initiator is
$H(t)+\eta_t$. Hence the maximum increases by at least $[\eta_t]_+$.
For each selected pair,
\[
 \E\!\left[\pos{\eta_t}\mid\F_t,I_t=i,R_t=r\right]
 =\sqrt{\frac{q_{ir}}{2\pi}}
 \ge\sqrt{\frac{q_{\min}}{2\pi}}.
\]
The responder is a maximiser with conditional probability at least $1/n$,
which gives \eqref{eq:effective-max-drift}. Taking expectations and summing
gives the stated bound on $\E H(t)$.

For any $A>0$, fresh pair-conditional Gaussian noise gives
\[
 \Pp(\eta_t>A\mid\F_t,I_t=i,R_t=r)
 \ge\Pp(N(0,q_{\min})>A)>0.
\]
Thus the event that a maximising responder is selected and $\eta_t>A$ has
a conditional probability bounded below uniformly in time. By
\lemref{lem:conditional-occurrence}, it occurs infinitely often with probability 1. Each occurrence raises $H$ by more than $A$ and leaves the
unchanged responder more than $A$ below the updated initiator. As in
\appref{app:right-align-plain}, a countable intersection over positive
integer $A$ proves $H(t)\to+\infty$ and $\limsup_tR(t)=\infty$ with probability 1.
\end{proof}

\subsection{Mean-square disagreement under joint noise}
\label{app:joint-ms-rate}

\begin{lemma}[Joint mean-square disagreement]
\label{lem:joint-ms-rate}
Under the assumptions of \thmref{thm:joint-relative},
\[
 \E V(t)=O(t^{-p}).
\]
Consequently,
\[
 \bigl\|(x_i(t)-x_j(t))-(p_i-p_j)\bigr\|_2=O(t^{-p/2})
\]
for every pair $i,j$.
\end{lemma}

\begin{proof}
Fix $0<\delta<p-1/2$ and let
\[
 G_t=\left\{\max_i|c_i(t)-t/n|\le t^{1/2+\delta}\right\}.
\]
On $G_t$, \lemref{lem:gain-control} gives
\[
 \gmin(t)=n^pt^{-p}(1+O(t^{-1/2+\delta})),\qquad
 \gmax(t)^2=n^{2p}t^{-2p}(1+O(t^{-1/2+\delta})).
\]
Substitution into \eqref{eq:joint-RS} yields
\[
 \E[V(t+1)\mid\F_t]
 \le \bigl(1-a_{n,p}^{\rm loc}t^{-p}+o(t^{-p})\bigr)V(t)
      +O(t^{-2p})
\]
on $G_t$, where $a_{n,p}^{\rm loc}=2n^p/(n-1)$ and $a_{n,1}^{\rm loc}>1$. The complement of $G_t$ has probability at most $2ne^{-2t^{2\delta}}$. The uniform conditional fourth-moment bound gives the polynomial moment envelope of \lemref{lem:moment-envelope}, so the contribution of $G_t^c$ to the unconditional recursion is superpolynomially small. Hence $u_t=\E V(t)$ satisfies the recursion in \lemref{lem:rate-comparison}, and $u_t=O(t^{-p})$.

Finally,
\[
 ((x_i-x_j)-(p_i-p_j))^2=(z_i-z_j)^2\le2V,
\]
which gives the relative-coordinate bound.
\end{proof}

\subsection{Common-translation rate for the joint model}
\label{app:joint-rate-proof}

\begin{proof}[Proof of \thmref{thm:joint-rate}]
Define
\[
 \bar q_i=\frac1{n-1}\sum_{r\ne i}q_{ir},\qquad
 S_p(\ell)=\sum_{m=\ell}^{\infty}(m+1)^{-2p}.
\]
Then $n^{-1}\sum_i\bar q_i=\bar q$. Conditional on the entire pair
schedule, the effective errors are independent centred Gaussians and the
noise-tail variance at $\tau=\tau_k^{\rm loc}$ is
\[
 \frac1{n^2}\sum_{t\ge\tau}\gamma_{I_t}(t)^2q_{I_tR_t}.
\]
Every agent has initiated at least $k$ times at $\tau$, so this is bounded by
\[
 \frac{q_{\max}}{n^2}\sum_i S_p(c_i(\tau))
 \le\frac{q_{\max}}n S_p(k),
\]
proving \eqref{eq:joint-schedule-bound}. The same bound at $k=0$ gives
$L^2$ convergence of $\mathcal Z$.

For the sharper unconditional constant, condition first on the entire
initiator sequence. The time $\tau$ and the starting counters $c_i(\tau)$
are functions only of that sequence. Conditional on an initiation by $i$,
the responder is uniform on the other agents; averaging over responders
therefore replaces $q_{ir}$ by $\bar q_i$. Orthogonality and the tower
property yield the exact identity
\begin{equation}
 \bigl\|\mathcal Z(\infty)-\mathcal Z(\tau)\bigr\|_2^2
 =\frac1{n^2}\E\!\left[\sum_i\bar q_i S_p(c_i(\tau))\right].
 \label{eq:joint-tail-identity}
\end{equation}
Counter concentration gives $\tau_k^{\rm loc}\sim nk$ and
$c_i(\tau_k^{\rm loc})/k\to1$ with probability 1 for every $i$. Also
$c_i(\tau_k^{\rm loc})\ge k$. By \eqref{eq:ptail},
\[
 \frac{S_p(c_i(\tau_k^{\rm loc}))}{k^{1-2p}}
 \longrightarrow\frac1{2p-1}
 \quad\text{with probability 1},
\]
and these ratios are bounded for $k\ge1$ by a deterministic constant,
since $S_p(c_i(\tau_k^{\rm loc}))\le S_p(k)$. Dominated convergence in
\eqref{eq:joint-tail-identity} proves \eqref{eq:joint-noise-tail}.
When $\bar q=0$, all effective errors vanish and both sides are zero.

For the remaining terms, use the mean decomposition
\[
 C_\infty-M(\tau)
 =\sum_{t\ge\tau}\mu_t+\sum_{t\ge\tau}U_t
  +\mathcal Z(\infty)-\mathcal Z(\tau),
\]
where $\mu_t$ is the predictable gain-mismatch drift and $U_t$ is the
pair-selection martingale difference from
\appref{app:signed-convergence-proof}. The bounds
\[
 |\mu_t|\le C_n\Dg(t)\sqrt{V(t)},\qquad
 \E[U_t^2\mid\F_t]\le C_n\gmax(t)^2V(t)
\]
are unaffected by the source of the noise. The Gaussian fourth moments
are bounded by $3q_{\max}^2$, and \thmref{thm:joint-relative} gives
$\E V(t)=O(t^{-p})$. Splitting over the good counter event and its
superpolynomially small complement as in \appref{app:signed-rate-proof}
therefore yields
\[
 \|\mu_t\|_2=O(t^{-3p/2-1/2+\delta}),\qquad
 \E[U_t^2]=O(t^{-3p}).
\]
Minkowski's inequality for the first series and orthogonality for the
second give
\[
 \left\|\sum_{t\ge\tau_k^{\rm loc}}\mu_t\right\|_2
 =O(k^{1/2-3p/2+\delta}),\qquad
 \left\|\sum_{t\ge\tau_k^{\rm loc}}U_t\right\|_2
 =O(k^{1/2-3p/2}).
\]
Here $\tau_k^{\rm loc}\ge k$, and
$\1_{\{\tau_k^{\rm loc}\le t\}}$ is $\F_t$-measurable, so the stopped-tail
martingale increments remain orthogonal. Combining these bounds with
\eqref{eq:joint-noise-tail} proves \eqref{eq:joint-common-tail}. If
$\bar q>0$, the correction terms are $o(k^{1/2-p})$ because $\delta<p/2$;
the reverse triangle inequality proves the stated asymptotic for the full
common-translation tail. No unconditional Gaussian law for the full
translation limit is asserted.
\end{proof}

\subsection{Examples, correlation, and model scope}
\label{app:joint-examples-scope}

\subsubsection{An explicit model with both variances increasing with distance}
\label{subsec:joint-example}

A simple illustrative choice is
\begin{equation}
 s_c^2(d)=\sigma_{c,0}^2+\alpha_c d,\qquad
 s_m^2(d)=\sigma_{m,0}^2+\alpha_m d,
 \qquad \alpha_c,\alpha_m>0,
 \label{eq:joint-affine}
\end{equation}
where the two baseline variances are nonnegative. Each error source now
has its own distance-sensitive term. This affine law is an example, not a
universal physical calibration; the theorems also cover different nonlinear
variance functions for communication and sensing.

Put $\sigma_0^2=\sigma_{c,0}^2+\sigma_{m,0}^2$,
$\alpha=\alpha_c+\alpha_m$, and
\begin{equation}
 \bar d=\frac1{n(n-1)}\sum_{i\ne r}d_{ir}.
 \label{eq:mean-distance}
\end{equation}
Then $q_{ir}=\sigma_0^2+\alpha d_{ir}$, so
\begin{equation}
 \begin{aligned}
  q_{\min}&=\sigma_0^2+\alpha d_{\min},\qquad
  q_{\max}=\sigma_0^2+\alpha d_{\max},\\
  \bar q&=\sigma_0^2+\alpha\bar d.
 \end{aligned}
 \label{eq:both-distance-constants}
\end{equation}
For this example, \RightAlign{} has conditional drift at least
\begin{equation}
 \frac{\sqrt{\sigma_0^2+\alpha d_{\min}}}{n\sqrt{2\pi}},
 \label{eq:both-max-drift}
\end{equation}
whereas \SignedUpdate{} converges and has common-translation tail
\begin{equation}
 \bigl\|C_\infty-M(\tau_k^{\rm loc})\bigr\|_2
 =\sqrt{\frac{\sigma_0^2+\alpha\bar d}{n(2p-1)}}\,
 k^{1/2-p}(1+o(1)).
 \label{eq:both-common-tail}
\end{equation}
Both distance-sensitive noise contributions therefore appear explicitly in
the leading constant. The communication-only and sensing-only cases are
listed separately in Appendix~\ref{app:single-source}.

\subsubsection{Correlated errors and scope of the model}
\label{subsec:joint-scope}

Independence between the two sources is not needed for signed convergence.
If their conditional means are zero and their conditional covariance is
$\kappa_{ir}$, then
\begin{equation}
 \begin{aligned}
  q_{ir}
  &=\Var(\zeta_t-\varepsilon_t\mid\F_t,I_t=i,R_t=r)\\
  &=s_c^2(d_{ir})+s_m^2(d_{ir})-2\kappa_{ir}.
 \end{aligned}
 \label{eq:correlated-effective-var}
\end{equation}
For example, a distance-dependent correlation coefficient $\rho(d)$ gives
$\kappa_{ir}=\rho(d_{ir})s_c(d_{ir})s_m(d_{ir})$, where
$|\rho(d)|\le1$. \thmref{thm:effective-signed} applies whenever
\eqref{eq:effective-assumptions} holds. If the errors are jointly Gaussian
with a fixed pair-dependent covariance and are fresh across interactions,
\thmref{thm:joint-rate} also applies with the resulting $\bar q$.
\obsref{obs:joint-max} assumes $q_{\min}>0$: correlated sources can
cancel in the effective difference, so nonzero component variances alone
do not imply this condition.

The distance-dependent formulation is a model of additive label and
measurement errors, not a claim that every communication channel or sensor
obeys such a law. If the conditional means are $b_c(d_{ir})$ and
$b_m(d_{ir})$, the effective bias is $b_c(d_{ir})-b_m(d_{ir})$; without
conditional centring, the martingale argument does not apply. Packet loss,
quantisation, clipping, and adversarial corruption require separate
assumptions. The scheduler remains uniform: distance affects the error
laws, not the probability that a pair interacts.

All guarantees concern fixed finite geometries. When the geometry or its
diameter changes with $n$, the constants may change; the stated exponents
are not uniform-in-$n$ bounds. For more general noise levels, a natural
condition to investigate is
\begin{equation}
 \sum_t\E\!\left[\gamma_{I_t}(t)^2\eta_t^2\mid\F_t\right]<\infty
 \quad\text{with probability 1}.
 \label{eq:generalvariance}
\end{equation}
This controls accumulated noise variance but does not by itself establish
tracking for moving agents, whose motion introduces additional terms into
the translation-error recursion.

\subsection{Single-source specialisations}
\label{app:single-source}

Communication-only and measurement-only models follow by setting the other
variance profile to zero.

\subsubsection{Communication errors only}
\label{subsec:communication-only}

Set $s_m^2(d)\equiv0$. Then $\eta_t=\zeta_t$ and
$q_{ir}=s_c^2(d_{ir})$, so the communication error may itself remain
distance-dependent. If instead $s_c^2(d)\equiv\sigma_c^2>0$, the model is
\begin{equation}
 \widetilde x_r=x_r+\zeta_t,\qquad \zeta_t\sim N(0,\sigma_c^2).
 \label{eq:scenario-comm}
\end{equation}
With fresh independent errors, \obsref{obs:joint-max} gives the drift
lower bound $\sigma_c/(n\sqrt{2\pi})$, while
\thmref{thm:effective-signed} gives signed convergence. This is the
localisation counterpart of fixed-law communication noise in
\secref{sec:averaging}.

\subsubsection{Measurement errors only}
\label{subsec:sensing-only}
\label{app:max3-proof}
\label{app:signed3-proof}

Set $s_c^2(d)\equiv0$. Then $\eta_t=-\varepsilon_t$ and
$q_{ir}=s_m^2(d_{ir})$. The constant choice
$s_m^2(d)\equiv\sigma^2$ recovers \secref{sec:localisation}. A
normalised distance-dependent example is
\begin{equation}
 \varepsilon_t\mid(\F_t,I_t=i,R_t=r)\sim N(0,d_{ir}),
 \qquad d_{ir}=|p_i-p_r|.
 \label{eq:scenario-distance}
\end{equation}
Here the proportionality coefficient has been set to one in the chosen
units. Then $q_{\min}=d_{\min}$, $q_{\max}=d_{\max}$, and
$\bar q=\bar d$, giving the following two consequences.

Setting $q_{ir}=d_{ir}$ in \obsref{obs:joint-max} gives
\begin{equation}
 \E[H(t+1)-H(t)\mid\F_t]
 \ge\frac{\sqrt{d_{\min}}}{n\sqrt{2\pi}},
 \label{eq:max3drift}
\end{equation}
and the same divergence of $H(t)$ holding with probability 1 and recurrent unbounded
relative-coordinate excursions.

\begin{samepage}
\begin{corollary}[Distance-dependent sensing: signed convergence]
\label{thm:signed3}
Under \eqref{eq:scenario-distance}, uniform independent ordered
interactions, and local gains \eqref{eq:gain} with $1/2<p\le1$,
\begin{equation}
 x_i(t)\longrightarrow p_i+C_\infty\qquad\text{for every }i
 \quad\text{with probability 1},
 \label{eq:loc3limit}
\end{equation}
and $\E V(t)=O(t^{-p})$. The noise-martingale tail satisfies
\begin{equation}
 \bigl\|\mathcal Z(\infty)-\mathcal Z(\tau_k^{\rm loc})\bigr\|_2
 =\sqrt{\frac{\bar d}{n(2p-1)}}\,k^{1/2-p}(1+o(1)).
 \label{eq:loc3meanrate}
\end{equation}
A schedule-conditional variance bound uses $d_{\max}$ instead of $\bar d$.
\end{corollary}

\begin{proof}
Apply \thmref{thm:effective-signed}, \thmref{thm:joint-relative}, and
\thmref{thm:joint-rate} with
$q_{ir}=d_{ir}$. The full common-translation tail has the same leading
asymptotic because the other terms are of smaller order, as shown in
\appref{app:joint-rate-proof}.
\end{proof}
\end{samepage}

\subsubsection{Fixed communication noise with distance-dependent sensing}

Set
$s_c^2(d)\equiv\sigma_c^2$ and $s_m^2(d)=d$. Then
$q_{ir}=\sigma_c^2+d_{ir}$ and $\bar q=\sigma_c^2+\bar d$.
This special case has constant communication variance; the affine model
\eqref{eq:joint-affine} allows both variances to increase with distance.

\clearpage

\begin{center}
    \includegraphics[scale=0.35]{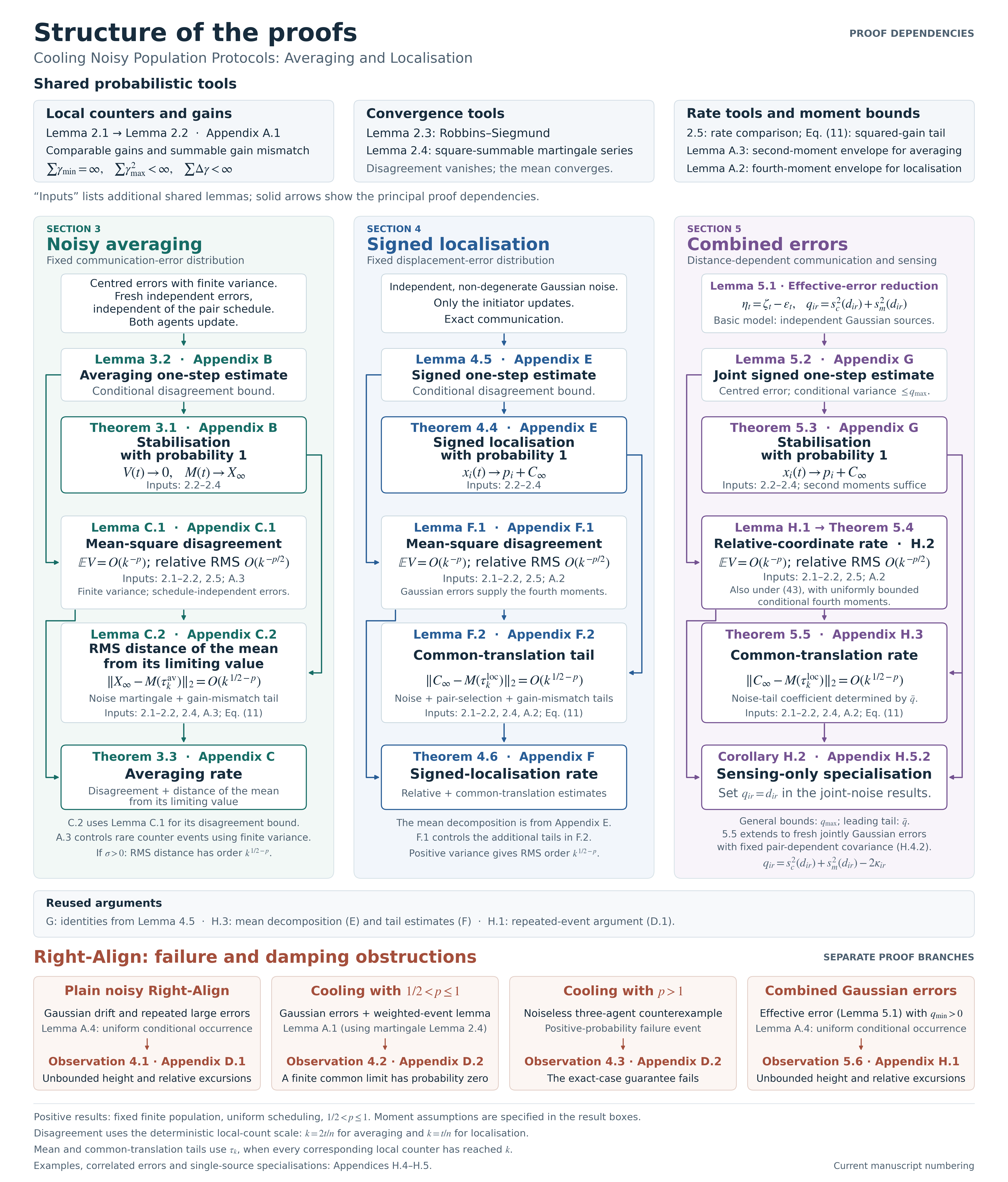}
\end{center}

\end{document}